\documentclass[12pt]{article}

\usepackage{graphicx}
\usepackage{alltt}
\usepackage{amsmath}
\usepackage{amssymb}
\usepackage{hyperref}
\usepackage{color}

\newcommand{\stl}{\rm\scriptstyle STL}
\newcommand{\tl}{\rm\scriptstyle TL}

\begin{document} 

\newcounter{Theorems}
\setcounter{Theorems}{0}

\newcounter{Definitions}
\setcounter{Definitions}{0}

\def\amklink#1#2{\href{http://andreimikhailov.com/#1}{\textcolor{blue}{\bf #2}}}

\begin{titlepage}
\begin{flushright}

\end{flushright}

\begin{center}
{\Large\bf $ $ \\ $ $ \\
A minimalistic pure spinor sigma-model in AdS
}\\
\bigskip\bigskip\bigskip
{\large Andrei Mikhailov${}^{\dag}$}
\\
\bigskip\bigskip
{\it Instituto de F\'{i}sica Te\'orica, Universidade Estadual Paulista\\
R. Dr. Bento Teobaldo Ferraz 271, 
Bloco II -- Barra Funda\\
CEP:01140-070 -- S\~{a}o Paulo, Brasil\\
}

\vskip 1cm
\end{center}

\begin{abstract}
The $b$-ghost of the pure spinor formalism in a general curved background is not holomorphic.
For such theories, the construction of the string measure requires the knowledge of the action
of diffeomorphisms on the BV phase space. We construct such an action for the pure spinor 
sigma-model in $AdS_5\times S^5$. From the point of view of the BV formalism, this sigma-model 
belongs to the class of theories where the expansion of the Master Action in antifields 
terminates at the quadratic order. We show that it can be reduced to a simpler degenerate 
sigma-model, preserving the AdS symmetries. We construct the action of the algebra of worldsheet 
vector fields on the BV phase space of this minimalistic sigma-model, and explain how to lift it 
to the original model.
\end{abstract}

\vfill
{\renewcommand{\arraystretch}{0.8}%
\begin{tabular}{rl}
${}^\dag\!\!\!\!$ 
& 
\footnotesize{on leave from Institute for Theoretical and 
Experimental Physics,}
\\    
&
\footnotesize{ul. Bol. Cheremushkinskaya, 25, 
Moscow 117259, Russia}
\\
\end{tabular}
}

\end{titlepage}

\tableofcontents 

\section{Introduction}
The $b$-ghost of the pure spinor formalism in a general curved background is only holomorphic
up to a $Q$-exact expression \cite{Berkovits:2010zz}. The construction of the string measure for such theories
was suggested in \cite{Mikhailov:2016myt,Mikhailov:2016rkp}. It requires the knowledge of the action of the group of worldsheet 
diffeomorphisms on the BV phase space. For a vector  field $\xi$ on the worldsheet (= infinitesimal
diffeomorphism) let $\Phi_{\xi}$ be the BV Hamiltonian generating the action of $\xi$ on the BV phase space.
Then, the string measure is, schematically:
\begin{equation}\label{StringMeasure}
\exp\left(S_{\rm BV} + \sigma + \Phi_F\right)
\end{equation}
where:
\begin{itemize}
\item $S_{\rm BV}$ is the worldsheet Master Action
\item $\sigma$ is the generating function of the variations of the Lagrangian submanifold (for
   the standard choice of the family, this is just the usual $\int \mu^z_{\bar{z}} b_{zz} + \mu^{\bar{z}}_{z} b_{\bar{z}\bar{z}}$)
\item $F$ is the curvature of the connection on the equivalence class of worldsheet theories, 
   considered as a principal bundle over the space of theories modulo diffeomorphisms
\end{itemize}
It is not completely trivial to construct $\Phi_{\xi}$ for the pure spinor superstring in AdS. One of the 
complications is the somewhat unusual form of the pure spinor part of the action. Schematically:
\begin{equation}\label{PureSpinorTerms}
S_{\lambda w}=\int \;w_{L+}(\partial_-+A_-)\lambda_L \;+\; w_{R-}(\partial_+ + A_+)\lambda_R \;+\; Sw_{L+}\lambda_Lw_{R-}\lambda_R
\end{equation}
where $S$ is a linear combination of Ramond-Ramond field strengths. Notice that the conjugate
momenta $w_L$ and $w_R$ only enter through their $(1,0)$ and $(0,1)$ component, respectively.
We can try to integrate out $w$, ending up with a ``standard'' kinetic term for ghosts:
\begin{equation}
{(\partial_- + A_-)\lambda_L\;(\partial_+ + A_+)\lambda_R\over S\lambda_L\lambda_R}
\end{equation}
Notice that $S$ landed in the denominator. It would seem that the theory depends quite 
irregularly on the Ramond-Ramond field, but this is not true. All physics sits at $\lambda=0$, 
and the $w\lambda w\lambda$ term is in some sense subleading. 

In this paper we will show, closely following \cite{Berkovits:2008ga,Tonin:2013uec}, that the pure spinor terms (\ref{PureSpinorTerms}) can actually 
be removed by reduction to a smaller BV phase space, keeping intact all the symmetries of 
$AdS_5\times S^5$. The resulting action is degenerate, and therefore can not be 
immediately used for quantization. On the other hand, it is simpler than the original action. 
In particular, the action of worldsheet diffeomorphisms in this reduced BV phase space is rather
transparent, although the explicit expression Eq. (\ref{MinimalPhi}) is somewhat involved. We then explain 
how to lift this action to an action on some quantizable theory which is basically the same
as the original pure spinor sigma-model of \cite{Berkovits:2000fe}. 

For the case of flat spacetime, the formal expressions are  somewhat more complicated. 
The construction of the action of diffeomorphisms is a work in progress with Renann Lipinski \cite{WithRenann}.

\paragraph     {Formal application of BV formalism}
Here, as in \cite{Mikhailov:2016rkp}, we formally apply the formalism of odd symplectic manifolds in the
infinite-dimensional case (the field space of two-dimensional sigma-models). This should be
proven in perturbation theory, but in this paper we restrict ourselves with purely formal
manipulations. We believe that supersymmetry will play crucial role in controlling
quantum anomalies; therefore it is important that our constructions preserve supersymmetries
(see Section \ref{sec:FormulationOfTheProblem}).

\paragraph     {Plan of the paper}
We begin in Section \ref{sec:BV} with the general discussion of the reduction procedure when a BV Master 
Action is a quadratic-linear functional of antifields. In Section \ref{sec:ActionWithoutW} we apply this to the case 
of pure spinor superstring in $AdS_5\times S^5$. In Sections \ref{sec:diffeomorphisms} we construct the action 
of diffeomorphisms in the minimalistic sigma-model. Then in Section \ref{sec:Regularization} we construct the action 
of diffeomorphisms on the BV phase space of the non-degenerate theory, which is essentially
equivalent (quasiisomorphic) to the original sigma-model. Sections \ref{sec:TakingApart} and
\ref{sec:Generalization} contain summary and generalizations, and Section \ref{sec:OpenProblems} open problems.

\section{Master Actions quadratic-linear in antifields}\label{sec:BV}
Suppose that the BV phase space is an odd cotangent bundle, {\it i.e.} is of the form $\Pi T^*N$
for some supermanifold $N$ (the ``field space''). If $\phi^a$ are coordinates on $N$, then
$\phi^{\star}_a$ are coordinates on $\Pi T^*N$, and  ``$\Pi$'' means that the statistics of $\phi^{\star}_a$ is opposite
to the statistics of $\phi^a$. There is an odd Poisson bracket (the ``BV bracket''):
\begin{equation}
\{\phi^{\star}_a,\phi^b\} = \delta_a^b
\end{equation}
This bracket is geometrically well-defined, in a sense that the bracket of two functions
$\{F,G\}$ is actually independent of how the coordinates $\phi^a$ on $N$ are choosen. Equivalently,
there is an
\amklink{math/bv/BV-formalism/Odd\_symplectic\_manifolds.html}{odd symplectic form}
(which, as any differential form, can be considered a function
on $\Pi T(\Pi T^*N)$):
\begin{equation}\label{OmegaBV}
\omega_{\rm BV} = \sum_a (-1)^{\bar{a}}d\phi^a\,d\phi^{\star}_a
\end{equation}
(As a slight overuse of Einstein notations, we will omit the summation sign $\Sigma_a$ in such cases.)
Suppose that the Master Action is of the form: 
\begin{equation}\label{OriginalAction}
S_{\rm BV} \;=\; S_{\rm cl}(\phi) + Q^a(\phi)\phi^{\star}_a + {1\over 2}\phi^{\star}_a\pi^{ab}(\phi)\phi^{\star}_b
\end{equation}
(writing $\phi^{\star}_a\pi^{ab}(\phi)\phi^{\star}_b$ rather than $\pi^{ab}(\phi)\phi^{\star}_a\phi^{\star}_b$ simplifies some signs later).

We will assume that $S_{\rm BV}$ satisfies the {\em classical Master Equation}:
\begin{equation}\label{ClassicalMasterEquation}
\{S_{\rm BV}, S_{\rm BV} \}  = 0
\end{equation}
If $N$ is purely even, we can think of functions on $\Pi T^*N$
as polyvector fields on $N$. For example, $Q^a(\phi)\phi^{\star}_a$ corresponds to the vector field
$Q^a(\phi){\partial\over\partial\phi^a}$, and ${1\over 2}\pi^{ab}(\phi)\phi^{\star}_a\phi^{\star}_b$ corresponds to a Poisson bivector $\pi^{ab}(\phi){\partial\over\partial\phi^a}\wedge{\partial\over\partial\phi^b}$.
The odd Poisson bracket $\{\_\,,\_\}$ corresponds to the Schouten bracket of polyvector fields.

If $N$ is a {\em super}-manifold, then this polyvector picture does not seem to be very
illuminating. However, one can still apply the intuition of Hamiltonian mechanics.
The linear function $Q=Q^a(\phi)\phi^{\star}_a$ still defines a vector field; the derivative of a function
$f\in C^{\infty}(N)$ along it is: $\left\{Q^a(\phi)\phi^{\star}_a\,,f(\phi)\right\} = Q^a\partial_af$. The quadratic function
$\pi = {1\over 2}\phi^{\star}_a\pi^{ab}(\phi)\phi^{\star}_b$ still defines a map from functions on $N$ to vector fields on $N$:
\begin{equation}
f \mapsto \{\pi, f\}
\end{equation}
The Master Equation  (\ref{ClassicalMasterEquation}) implies, order by order in expansion in $\phi^{\star}$:
\begin{align}
  \{Q,S_{\rm cl}\} \;=\; & 0
  \\
  \{Q,Q\}  + 2\{\pi, S_{\rm cl}\} \;=\;&  0
\label{QOnlyNilpotentOnShell}\\    
\{Q,\pi\} \;=\;&0
\label{SchoutenQPi}\\   
\{\pi,\pi\} \;=\;& 0
\label{Jacobi}
\end{align}
It follows from Eq. (\ref{Jacobi}) that vector fields of the form $\{\pi,f\}$, $f\in C^{\infty}(N)$,
form a closed subalgebra in the algebra of vector fields. They are all tangent to a family of
submanifolds of $N$ which can be called ``symplectic leaves of $\pi$''. As a slight abuse of notations,
the letter $Q$ will denote both the BRST transformation $Q^a\partial_a$ and the function $Q^a\phi^{\star}_a$ on $\Pi T^*N$.
Eq. (\ref{QOnlyNilpotentOnShell}) says that generally speaking the BRST operator $Q$ is only nilpotent on-shell \cite{Berkovits:2007rj}.

We will show that under some conditions, this theory can be reduced to a simpler theory which 
has BRST operator nilpotent off-shell (and therefore its Master Action has no 
quadratic terms $\phi^{\star}\phi^{\star}$). 

\paragraph     {The case when $\pi$ is non-degenerate}
Let us first consider the case when the Poisson bivector $\pi^{ab}$ is nondegenerate. Eq. (\ref{SchoutenQPi}) 
implies that an odd function $\psi\in \mbox{Fun}(N)$ locally exists, such that $Q = \{\pi,\psi\}$. 
Suppose that $\psi$ is also defined globally. Let us consider the canonical transformation of
the Darboux coordinates generated by $\psi$:
\begin{align}
(\phi,\phi^{\star}) \;\to\; & (\tilde{\phi},\tilde{\phi}^{\star})
\nonumber\\   
\phi^a \;=\; & \tilde{\phi}^a
\label{CanonicalTransformations}\\     
\phi^{\star}_a \;=\; & \tilde{\phi}^{\star}_a + {\partial\over\partial\tilde{\phi}^a}\psi(\tilde{\phi})
\nonumber
\end{align}
More geometrically: $\tilde{\phi}$ and $\tilde{\phi}^{\star}$ (functions on $\Pi T^*N$) are pullbacks of $\phi$ and $\phi^{\star}$ by 
the flux of the Hamiltonian vector field $\{\psi,\_\}$ by the time $1$. (The flux integrates to
Eqs. (\ref{CanonicalTransformations}) because $\psi$ only depends on $\phi$, 
and therefore the velocity of $\phi^{\star}$ is $\phi^{\star}$-independent.)

In the new coordinates:
\begin{align}
S_{\rm} \;=\; & \widetilde{S}_{\rm cl} + {1\over 2}\phi^{\star}_a\pi^{ab}(\phi)\phi^{\star}_b
\label{SplitAction}
\\  
\mbox{\tt\small where } &
\widetilde{S}_{\rm cl}\;=\; S_{\rm cl} + 
{1\over 2} \partial_a\psi\pi^{ab}\partial_b\psi 
\end{align}
The $\phi^{\star}$-linear term is gone! The Master Equation implies that $\{\widetilde{S}_{\rm cl},\pi\} = 0$. 
Since we assumed that $\pi$ is nondegenerate, this implies:
\begin{equation}
\widetilde{S}_{\rm cl}\;=\; \mbox{const}
\end{equation}

\paragraph     {The case of degenerate $\pi$}
We are actually interested in the case when $\pi$ is degenerate. Let ${\cal P}\subset TN$ be the 
distribution tangent to symplectic leaves of $\pi$:
\begin{equation}
{\cal P} =\mbox{im}\,\pi \subset TN
\end{equation}
This distribution is integrable because $\pi$ satisfies the Jacobi identity.
We also assume that $Q$ is transverse to $\cal P$:
\begin{equation}
Q\notin {\cal P}
\end{equation}
 Let us also consider 
the distribution ${\cal P} + {\cal Q}$ which is generated by elements of ${\cal P}$ and by $Q$.
Eqs. (\ref{SchoutenQPi}) and (\ref{Jacobi}) imply that ${\cal P} + {\cal Q}$  is also integrable. Let us {\em assume} the
existence of a 2-form\footnote{This $\omega$ is {\em even}; it should not be confused with the odd symplectic form of $\Pi T^*N$.} $\omega$ on each integrable surface\footnote{It is enough to define $\omega$ on each integrable surface of ${\cal P} + {\cal Q}$; it does
  not have to be defined on the whole $N$.} of ${\cal P} + {\cal Q}$ and a function $\psi\in \mbox{Fun}(N)$ which satisfy:
\begin{align}
\pi\omega\pi \;=\; & \pi
\label{PiOmegaPi}\\      
\omega\pi\omega \;=\; &  \omega
\label{OmegaPiOmega}\\    
d\omega|_{{\cal P} + {\cal Q}} \;=\; & 0
\label{OmegaIsClosed}
\\   
\left.(\iota_Q\omega - d\psi)\right|_{{\cal P}} \;=\; & 0
\label{IotaQOmega}
\end{align}
where $\pi\omega\pi$ and $\omega\pi\omega$ are defined as follows:
\begin{align}
  \phi^{\star}_a\left(\pi\omega\pi\right)^{ab}\phi^{\star}_b \;=\;
  & \phi^{\star}_{a}\pi^{aa'} \omega_{a'b'}\pi^{b'b}\phi^{\star}_b
  \\
  d\phi^a \left(\omega\pi\omega\right)_{ab} d\phi^b \;=\;
  & d\phi^a \omega_{aa'}\pi^{a'b'}\omega_{b'b} d\phi^b
\end{align}
Existence of $\psi$ satisfying Eq. (\ref{IotaQOmega}) locally follows from Eqs. (\ref{PiOmegaPi}) and (\ref{OmegaIsClosed}), because they 
imply $\left.d(\iota_Q\omega)\right|_{\cal P} = 0$. But we also require this $\psi$ to be a {\em globally} well-defined 
function on $N$. Contracting $\iota_Q\omega - d\psi$ with $\pi\omega$ we find that: 
\begin{equation}\label{NewQIsInKerOmega}
Q - \{\pi,\psi\} \in \mbox{ker}\,\left(\omega|_{{\cal P} + {\cal Q}}\right)
\end{equation}
Let us define the new odd vector field:
\begin{equation}\label{NewBRSTOperator}
\widetilde{Q} = Q - \{\pi,\psi\}
\end{equation}
Eq. (\ref{OmegaIsClosed}) implies that $\mbox{ker}\,\left(\omega|_{{\cal P} + {\cal Q}}\right)$ is an integrable distribution inside an integral 
surface of ${\cal P} + {\cal Q}$. Therefore Eq. (\ref{NewQIsInKerOmega}) implies that $\widetilde{Q}^2$ is proportional to $\widetilde{Q}$, {\it i.e.} 
there exists a function $\zeta$ such that: $\widetilde{Q}^2=\zeta\widetilde{Q}$. In fact $\zeta=0$, since $\widetilde{Q}^2\in{\cal P}$ and $\widetilde{Q}\notin {\cal P}$. 
We conclude:
\begin{equation}
\widetilde{Q}^2 = 0
\end{equation}
Let us consider the canonical transformation (\ref{CanonicalTransformations}) of Darboux coordinates generated by $\psi$.
With these new Darboux coordinates:
\begin{align}\label{NewDarbouxCoordinates}
S_{\rm BV}\;=\;& S_{\rm cl} - {1\over 2}\omega(Q,Q) \;+\; 
\left(Q -  \{\pi,\psi\}\right)^a\tilde{\phi}^{\star}_a \;+\; 
{1\over 2}\tilde{\phi}^{\star}_a\pi^{ab}\tilde{\phi}^{\star}_b
\end{align}
Notice that the new ``classical action'':
\begin{equation}\label{SclIsConst}
\tilde{S}_{\rm cl} = S_{\rm cl} - {1\over 2}\omega(Q,Q)
\end{equation}
is automatically constant on symplectic leaves of $\pi$. Also, it follows that $\widetilde{Q}$ consistently 
defines an odd nilpotent vector field on the moduli space of symplectic leaves of $\pi$. 
These facts follow from $\{S_{\rm BV},S_{\rm BV}\}=0$. To summarize:
\begin{align}
S_{\rm BV} \;=\; & \tilde{S}_{\rm BV} + {1\over 2}\tilde{\phi}^{\star}_a\pi^{ab}(\phi)\tilde{\phi}^{\star}_b
\\    
\mbox{\tt\small where } & 
\tilde{S}_{\rm BV} = \tilde{S}_{\rm cl}(\chi) + \tilde{Q}(\chi)^m\chi^{\star}_m
\end{align}
where $\chi$ is coordinates on the space of symplectic leaves of $\pi$. We therefore constructed a new, 
simpler theory, on the space of symplectic leaves of $\pi$.

This theory can be interpreted as the result of
\amklink{math/bv/transfer/Partial_Integration.html}{integrating out} 
some antifields. More precisely, let us define a submanifold $N_0\subset N$ by picking one point
from each symplectic leaf. Fibers of the
\amklink{math/bv/BRST-formalism/Family_of_Lagrangian_submanifolds.html\#(part.\_.Conormal\_bundle)}{odd conormal bundle}\footnote{The fiber of the conormal bundle of $N_0\subset N$ at the point $\phi\in N_0$ consists of those elements of $T_{\phi}^*N$ which vanish on $T_{\phi}N_0\subset T_{\phi}N$.}
$\Pi T^* N_0$ are isotropic submanifolds in $\Pi T^*N$, and we can integrate them out as
described in \cite{Mikhailov:2016rkp}.
In this paper the coordinates in these fibers will be called $w^{\star}$ (and integrated out).

\paragraph     {Oversimplified example}
We will now illustrate the relation by a toy sigma-model (we will actually run the procedure
``in reverse''). Let $\Sigma$ be a two-dimensional worldsheet. Let us start with:
\begin{align}
S_{BV} \;=\; & S_{cl} +  \int_{\Sigma} \lambda \theta^{\star}
\end{align}
where $S_{\rm cl}$ does not not depend neither on the fermionic field $\theta^a$ nor on the bosonic field $\lambda^a$.
(It depends on some other fields $\phi^{\mu}$.) We postulate the odd symplectic form so that our fields
are
\amklink{math/bv/BV-formalism/Odd_symplectic_manifolds.html}{Darboux coordinates}
\cite{Mikhailov:2016rkp}, as in Eq.~(\ref{OmegaBV}):
\begin{equation}
   \omega_{\rm BV} \;=\; \int_{\Sigma}d\lambda^{\star} d\lambda - d\theta^{\star} d\theta +
   \sum\limits_{\mu\in \left\{{\tt other\atop fields}\right\}}(-1)^{\bar{\mu}}d\phi_{\mu}^{\star}d\phi^{\mu}
\end{equation}
This action is highly degenerate; the path integral  $\int [d\lambda][d\theta][d\phi]e^{S_{\rm cl}(\phi)}$ is undefined
(infinity from integrating over $\lambda$ times zero from integrating over $\theta$). To regularize $\infty\times 0$,
let us introduce a new field-antifield pair $w,w^{\star}$, where $w$ is a bosonic 1-form on the
worldsheet and $w^{\star}$ is a fermionic 1-form on the worldsheet:
\begin{align}
  & w \;=\; w_+dz + w_-d\overline{z}
  \\   
  & w^{\star} \;=\; w^{\star}_+dz + w^{\star}_- d\overline{z}
\end{align}
The total odd symplectic form is postulated as follows:
\begin{equation}
  \omega^{\rm tot}_{\rm BV} \;=\; \omega_{\rm BV} + \int_{\Sigma} dw^{\star}\wedge dw
\end{equation}
(where $d$ is the field space differential, {\em not} the worldsheet differential).
Let us add $(w^{\star})^2$ to the BV action:
\begin{equation}\label{OldBVAction}
S_{BV} = S_{cl} + \int \lambda \theta^{\star} + \int w^{\star}\wedge w^{\star}
\end{equation}
(Notice that this $\int w^{\star}\wedge w^{\star}$ does not involve the worldsheet metric.) This corresponds to:
\begin{equation}
   \omega = \int_{\Sigma} dw\wedge dw
\end{equation}
(again, $d$ is the field space differential, {\em not} the worldsheet differential). In this case
${\cal P}$ is the subspace of the tangent space generated by $\partial\over\partial w$, and $\cal Q$ is generated by $\lambda{\partial\over\partial\theta}$. 
Then, shift the Lagrangian submanifold by a gauge fermion:
\begin{equation}\label{ToyGaugeFermion}
\Psi = \int_{\Sigma} w\wedge d\theta
\end{equation}
This results in the new classical action:
\begin{align}
  S_{\rm cl}^{\rm new} \;=\;
  & S_{\rm cl} + \int_{\Sigma} w\wedge d\lambda + \int_{\Sigma} d\theta\wedge d\theta
    \label{SimplifiedOriginalAction}
  \\   
  S_{\rm BV} \;=\;
  & S_{\rm cl}^{\rm new} + \int \lambda\theta^{\star} + \int d\theta \,w^{\star} +
    \int w^{\star}\wedge w^{\star}
  \\
  Q^{\rm new} \;=\;
  & \lambda{\partial\over\partial\theta} + d\theta {\partial\over\partial w}
\end{align}
Here we have run the procedure of Section \ref{sec:BV} ``in reverse''. That is, Eq. (\ref{SimplifiedOriginalAction}) is an example
of the $S_{\rm cl}$ of Eq. (\ref{OriginalAction}), and Eq. (\ref{OldBVAction}) is an example of the ``split'' Eq. (\ref{SplitAction}). Notice that
$\pi$ is degenerate, as it does not involve $\partial\over\partial\theta$ and $\partial\over\partial\lambda$. Because of that, the $S_{\rm cl}$ of Eq. (\ref{OldBVAction}) is
not constant as in Eq. (\ref{SclIsConst}), but just independent of $w$. The vector field $\{\pi,\Psi\}$
is the $d\theta{\partial\over\partial w}$-part of $Q^{\rm new}$, as in Eq.~(\ref{NewQIsInKerOmega}).

This is, still, not a quantizable action (the kinetic term for $\theta$ is a total derivative).
One particular way of choosing a Lagrangian submanifold leading to quantizable action is
to treat $w_+$ and $w_-$ asymmetrically (pick a worldsheet complex structure),
see section on A-model in AKSZ\cite{Alexandrov:1995kv} and Section \ref{sec:ConormalBundle} of this paper. This requires more than one
flavour of $w$.

\section{Pure spinor superstring in $AdS_5\times S^5$}\label{sec:ActionWithoutW}
\subsection{Notations}
We follow the notations in \cite{Bedoya:2010qz}.
The superconformal algebra ${\bf g} = {\bf psu}(2,2|4)$ has ${\bf Z}_4$-grading:
\begin{equation}
   {\bf g} = 
   {\bf g}_{\bar{0}} + {\bf g}_{\bar{1}} + {\bf g}_{\bar{2}} + {\bf g}_{\bar{3}} 
\end{equation}
Bars over subindices are to remind that they are mod 4.
Geometrically, ${\bf g}_{\bar{2}}$ can be identified with the tangent space to the bosonic $AdS_5\times S^5$, 
which is the direct sum of the tangent space to $AdS_5$ and the tangent space to $S^5$:
\begin{equation}
   T(AdS_5\times S^5) = T(AdS_5) \oplus T(S^5)
\end{equation}
Therefore elements of ${\bf g}_{\bar{2}}$ are vectors from this tangent space. We can also consider
the tangent space to the full superspace $M$: \marginpar{$M$}
\begin{align}
& M = {\rm super}(AdS_5\times S^5) = {PSU(2,2|4)\over SO(1,4)\times SO(5)}
\label{SuperAdS}\\      
& T\left({PSU(2,2|4)\over SO(1,4)\times SO(5)}\right) = {\bf g}_{\bar{1}} \oplus {\bf g}_{\bar{2}} \oplus {\bf g}_{\bar{3}}
\end{align}
--- this is a direct sum of three vector bundles. 
We parametrize a point in $M$ by $g\in PSU(2,2|4)$ modulo the equivalence relation:
\marginpar{$g$}
\begin{equation}\label{CosetEquivalence}
g\simeq hg \;\;\mbox{ \tt\small for all }\;\; h\in SO(1,4)\times SO(5)
\end{equation} 
We are identifying representations of ${\bf g}_{\bar{0}}=\mbox{Lie}(SO(1,4)\times SO(5))$, such as ${\bf g}_{\bar{1}}$, ${\bf g}_{\bar{2}}$, ${\bf g}_{\bar{3}}$,
with the corresponding vector bundles over the coset space (\ref{SuperAdS}). In fact, the worldsheet field 
$\lambda_L$ takes values in the fibers of ${\bf g}_{\bar{3}}$ and $\lambda_R$ takes values in the fibers of ${\bf g}_{\bar{1}}$. The pure spinor
conditions define the cones $C_L$ and $C_R$:\marginpar{$C_L,C_R$}
\begin{align}
C_L\;:\;\{\lambda_L,\lambda_L\} \;=\; & 0
\\   
C_R\;:\;\{\lambda_R,\lambda_R\} \;=\; & 0
\end{align}
Here $\{\_,\_\}$ denotes the anticommutator (the Lie superalgebra operation) of elements of ${\bf g}$.
It should not be confused with neither the odd Poisson bracket, nor the even Poisson bracket
corresponding to $\pi^{ab}$ of Section \ref{sec:BV}.
Again, we identify $C_L$ and $C_R$ as bundles over super-AdS. (They are not {\em vector} bundles, 
because their fibers are cones and not linear spaces.) We will denote:
\marginpar{$PS\;AdS$}
\begin{equation}\label{PSAdS}
PS\;AdS_5\times S^5 \; = \; {C_L\times C_R\times PSU(2,2|4)\over SO(1,4)\times SO(5)}
\end{equation}
where the prefix PS on the LHS stands for ``Pure spinors'' (and on the RHS for ``Projective''
and ``Special'').

\vspace{10pt}
\noindent
In Appendix \ref{sec:Projector} we construct $PSU(2,2|4)$-invariant surjective maps of bundles 
(``projectors''):\marginpar{${\bf P}_{31}$}
\begin{align}
{\bf P}_{31}\;:\; & ({\bf g}_{\bar{3}}\times C_L) \rightarrow TC_L
\\    
{\bf P}_{13}\;:\; & ({\bf g}_{\bar{1}}\times C_R) \rightarrow TC_R 
\end{align}
They are rational functions of $\lambda_L$ and $\lambda_R$.

\subsection{Standard action}
The action of the AdS sigma-model has the following form \cite{Berkovits:2000fe}:
\begin{align}\label{ActionSplit}
   S_0 \;=\; 
   \int dz\; d\bar{z} \; \mbox{Str}\Big( &
      {1\over 2} J_{2+}J_{2-} + {3\over 4} J_{1+}J_{3-} + {1\over 4} J_{3+}J_{1-} \;+ 
\\    
& + 
      w_{1+}D_{0-}\lambda_3 + w_{3-}D_{0+}\lambda_1 - N_{0+}N_{0-}
   \Big) 
\nonumber
\end{align}
where $J_{n}$ are the ${\bf g}_{\bar{n}}$-components of $J=-dgg^{-1}=J_+ dz + J_- d\bar{z}$. We write $\lambda_3$ instead
of $\lambda_L$ and $\lambda_1$ instead of $\lambda_R$, just to highlight the ${\bf Z}_4$-grading.
\marginpar{{\tiny notations}\\ $\lambda_3=\lambda_L$\\  $\lambda_1=\lambda_R$}
(And also because neither $\lambda_L$ is strictly speaking left-moving, nor is $\lambda_R$ right-moving.)
The covariant derivative $D_{0\pm}$ is defined as follows:
\begin{equation}
D_{0\pm} = \partial_{\pm} + [J_{0\pm},\_]
\end{equation}
Since $\lambda_3$ and $\lambda_1$ both satisfy the pure spinor constraints, the corresponding conjugate 
momenta are defined up to ``gauge transformations'':
\begin{align}
   \delta_{v_2} w_{1+} \;&= [v_{2+}, \lambda_3]
   \label{GaugeTransformationOfWPlus}
   \\   
   \delta_{u_2} w_{3-} \;&= [u_{2-}, \lambda_1]
   \label{GaugeTransformationOfWMinus}
\end{align}
where $v_2$ and $u_2$ are arbitrary sections of the pullback to the worldsheet of ${\bf g}_{\bar{2}}$. The BRST 
transformations are defined up to gauge transformations corresponding to the equivalence 
relation (\ref{CosetEquivalence}). It is 
\amklink{slides/talk\_Perimeter/LiftOfQ.html}{possible to fix this ambiguity} so that:
\begin{align}
  & Q\lambda_L = Q\lambda_R \; = 0
    \label{QLambda}\\   
  & Qg \; = (\lambda_L + \lambda_R)g
    \label{Qg}\\  
  & Qw_{1+}\;= -J_{1+}\;,\;\;Q w_{3-} \;= -J_{3-}
    \label{Qw}
\end{align}
The first line in Eq. (\ref{ActionSplit})  is by itself {\em not} BRST invariant. Modulo total derivatives,
its BRST variation is:
\begin{align}
   Q&\; \int d\tau\; d\sigma \; \mbox{Str}\left(
      {1\over 2} J_{2+}J_{2-} + {3\over 4} J_{1+}J_{3-} + {1\over 4} J_{3+}J_{1-}
   \right)\;=
   \nonumber \\   
   =&\; \int d\tau\; d\sigma \; \mbox{Str}\left(
      - D_{0+}\lambda_1\;J_{3-} -  D_{0-}\lambda_3\;J_{1+}
   \right)
   \label{VariationOfTheMatterPart}
\end{align}
This  cancels with the BRST variation of the second line in Eq. (\ref{ActionSplit}).

\subsection{New action}
On the other hand, we observe that:
\begin{equation}
   Q\; \mbox{STr}\left(
      J_{1+}{\bf P}_{31}J_{3-}
   \right) = \mbox{STr}\left(
      -  D_{0+}\lambda_1 \; J_{3-}
      -  D_{0-}\lambda_3 \; J_{1+}
   \right)
\end{equation}
Notice that the projector drops out on the RHS because $D_{0\pm}\lambda$ is automatically tangent 
to the cone. Comparing this to (\ref{VariationOfTheMatterPart}) we see that the following expression:
\begin{equation}\label{NewAction}
   S_0' = \int\;d\tau\;d\sigma \; \mbox{STr}\left(
      {1\over 2}J_{2+}J_{2-} + {3\over 4}J_{1+}J_{3-} + {1\over 4}J_{3+}J_{1-} 
      - J_{1+}{\bf P}_{31}J_{3-}
   \right)
\end{equation}
is BRST invariant. It does not contain neither derivatives of pure spinors, nor their conjugate
momenta. 

\subsection{The $b$-ghost}\label{sec:B-ghost}
We define:
\begin{align}
b_{++} \;=\; & {\mbox{STr}\left((\{J_{3+},\lambda_3\} - \{J_{1+},\lambda_1\}) \overline{J_{2+}}\right)\over\mbox{STr}(\lambda_3\lambda_1)}\;=\;
{\mbox{Tr}\left((\{J_{3+},\lambda_3\} - \{J_{1+},\lambda_1\})J_{2+}\right)\over\mbox{STr}(\lambda_3\lambda_1)}
\\   
b_{--} \;=\; & \mbox{\tt\small same but with $+$ replaced with $-$}
\end{align}
(See Appendix \ref{sec:Projector} for notations. We use the fact that 
$\mbox{Str}(A_2\overline{B_2}) = \mbox{Str}(A_2B_2\Sigma) = \mbox{Tr}(A_2B_2)$.)
These expressions satisfy (Appendix \ref{sec:BRSTofB}): 
\begin{align}
& Qb_{++} = T_{++}\mbox{ \tt\small and } Qb_{--} = T_{--} 
\\   
\mbox{\tt\small where } 
& T_{++} = \mbox{Str}\left(
  {1\over 2}J_{2+}J_{2+} + J_{1+}({\bf 1} - {\bf P}_{31})J_{3+}
\right)
\nonumber\\  
& T_{--} = \mbox{Str}\left(
  {1\over 2}J_{2-}J_{2-} + J_{1-}({\bf 1} - {\bf P}_{31})J_{3-}
\right)
\nonumber
\end{align}
Notice that:
\begin{align}
  S_0' \;=\;&  S_0''  + Q B
  \label{GaugingAwayG}\\
  \mbox{\tt  where } &
  \nonumber\\
  B \;=\;& \int d\tau d\sigma 
      {\mbox{Tr}\Big((\{J_{3+},\lambda_3\} - \{J_{1+},\lambda_1\})J_{2-} + (+\leftrightarrow -)\Big)
      \over
      \mbox{STr}(\lambda_3\lambda_1)}
  \\
  S_0'' \;=\;
  & \int \mbox{STr}\left(
    J_1\wedge (1-{\bf P}_{31})J_3 - J_1\wedge {\bf P}_{31} J_3
    \right)
\end{align}
and $S_0''$ is diffeomorphism-invariant (and therefore degenerate!).
The BRST invariance of $S_0''$ can be verified explicitly as follows:
\begin{align}
  QS_0''\;=\;& \int \mbox{STr}\Big( [\lambda_3,J_2]\wedge J_3 - [\lambda_1,J_2]\wedge J_1
               -D_0\lambda_1\wedge J_3 + D_0\lambda_3\wedge J_1\Big)\;=
               \nonumber  \\
  \;=\;& \int d\;\mbox{STr}(\lambda_3J_1 - \lambda_1J_3)\;=\;0
\end{align}

\subsection{Gauge fixing $SO(1,4)\times SO(5)$}\label{sec:A0}
Consider  the action of the  BRST operator given by Eq (\ref{Qg}) on $g$. It is nilpotent only up to
the ${\bf g}_0$-gauge transformation by $\{\lambda_3,\lambda_1\}$. We have so far worked on the factorspace by gauge
transformations. This means that we think of the group element $g$ and pure spinors $\lambda$ as 
defined only modulo the gauge transformation: 
\begin{equation}\label{GaugeTransformationOfG}
(g,\lambda) \simeq (hg, h\lambda h^{-1})
\end{equation}
It turns out that the action of these gauge transformations on the BV phase space is somewhat
nontrivial, see Section \ref{sec:GluingCharts}. We will now just fix the gauge, postponing the discussion of gauge
transformations to Section \ref{sec:GluingCharts}. Let us
parametrize the group element $g\in PSU(2,2|4)$ by $u,x,\theta$:\marginpar{$x$ and $\theta$} 
\begin{equation}\label{ParametrizationOfG}
g = e^ue^{x+\theta}
\end{equation}
where $u\in {\bf g}_0$, $x\in {\bf g}_2$ and $\theta\in {\bf g}_3 + {\bf g}_1$, and impose the following gauge fixing condition:
\begin{equation}\label{GaugeCondition}
u=0
\end{equation}
Since Eq. (\ref{GaugeCondition}) does not contain derivatives, this gauge is ``ghostless'',
the Faddeev-Popov procedure is not needed\footnote{The Faddeev-Popov procedure in such cases leads to ghost
  action of the form $\int f(\phi)\bar{c}c$ where $f(\phi)$ is some function of the fields.
  Integration out $c$ and $\bar c$ leads to {\em local} expressions
  (in fact, proportional to $\delta(0)$)
  which are absorbed by counterterms. Similar topics were discussed in \cite{Apfeldorf:1994av,Kreimer:2012qu}}.
In this gauge fixed formalism, the BRST operator includes the gauge
fixing term ({\it cp.} Eqs. (\ref{QLambda}), (\ref{Qg}), (\ref{Qw})):\marginpar{$A_0$}
\begin{align}
  & Qg = (\lambda_3+ \lambda_1 + A_0)g
    \label{QgWithA0} \\    
  & Q\lambda_3 = [A_0,\lambda_3]\;,\;\;Q\lambda_1 = [A_0,\lambda_1]
    \label{QLambdaWithA0} \\
  & Qw_{1+}\;= -J_{1+}+ [A_0,w_{1+}]\;,\;\;Q w_{3-} \;= -J_{3-} + [A_0,w_{3-}]
    \label{QwWithA0}
\end{align}
where $A_0\in {\bf g}_{\bar 0}$ is some function of $\theta$, $\lambda$ and $x$, defined by Eqs. (\ref{QgWithA0}) and (\ref{GaugeCondition});
schematically $A_0 = \{\theta_L,\lambda_1\} + \{\theta_R,\lambda_3\} + \ldots$
This $A_0$ is usually called ``the compensating gauge transformation''. It automatically satisfies:
\begin{equation}
QA_0= -\{\lambda_3,\lambda_1\} + {1\over 2}[A_0,A_0]
\end{equation}
Gauge fixing is only possible locally in $AdS_5\times S^5$. In order for our constructions to work
globally, we will cover $AdS_5\times S^5$ with patches and gauge-fix over each patch. Then we have
to glue overlapping patches. We will explain how to do this in Section \ref{sec:GluingCharts}.

\subsection{In BV language}
We will now show that the difference between the original action and the action (\ref{NewAction}) can be
interpreted in the BV formalism as a particular case of the construction outlined in Section \ref{sec:BV}.

The BRST symmetry of the pure spinor superstring in $AdS_5\times S^5$ is nilpotent only
on-shell. More precisely, the only deviation from the nilpotence arises when we act
on the conjugate momenta of the pure spinors:
\begin{align}
   Q^2w_{1+} \;&= {\delta S_0\over \delta w_{3-}}
   \\  
   Q^2w_{3-} \;&= {\delta S_0\over \delta w_{1+}}
\end{align}
(while the action of $Q^2$ on the {\em matter} fields is zero even off-shell). 
This means that the BV Master Action contains a term quadratic in the antifields: 
\begin{align}
  S_{\rm BV} \;
  &= S_0 + \int (QZ^i)Z^{\star}_i + \int (Q\lambda)\lambda^{\star} + \int (Qw)w^{\star} +
  \nonumber \\   
  \;& + \int\mbox{Str}\;\left( w^{\star}_{1+}w^{\star}_{3-}\right)
      \label{StandardBV}
\end{align}
In this formula $Z$ and $Z^{\star}$ stand for matter fields ($x$ and $\theta$) and their antifields, and
$S_0$ is given by Eq. (\ref{ActionSplit}). The matter
fields $Z$ are essentially $x$ and $\theta$ where $J=-dg g^{-1}$ with $g=e^{x+\theta}$, $x\in{\bf g}_2$,
$\theta\in {\bf g}_3\oplus {\bf g}_1$:
\begin{equation}
   Z\;=\;\;x\mbox{ \small\tt and }\theta
\end{equation}
Their BRST transformation $QZ^i$ is read from Eq. (\ref{QgWithA0}). We observe that the action is of
the same type as described in Section \ref{sec:BV}. The Poisson bivector is:
\begin{equation}
\pi = \int\mbox{Str}\left({\partial\over\partial w_{1+}}\wedge {\partial\over\partial w_{3-}}\right)
\end{equation}
The 2-form $\omega$ discussed in Section \ref{sec:BV} can be choosen as follows:
\begin{equation}
\omega = \int\mbox{Str}\left(dw_{1+}\wedge {\bf P}_{31}dw_{3-}\right)
\end{equation}
The projector ${\bf P}_{31}$ is needed to make $\omega$ invariant with respect to the gauge
transformations (\ref{GaugeTransformationOfWPlus}) and (\ref{GaugeTransformationOfWMinus}). We take the following generating function $\psi$ satisfying
Eq. (\ref{IotaQOmega}):
\begin{equation}
   \psi = \int \mbox{Str}\;\left(
      w_{1+}{\bf P}_{31}J_{3-} + w_{3-}{\bf P}_{13}J_{1+} + w_{1+}[A_0,w_{3-}]
   \right)
\end{equation}
The new ``classical action'' $\widetilde{S}_{\rm cl}$ is given by Eq (\ref{NewAction}). (We will provide more details for a
slightly more general calculation in Section \ref{sec:Regularization}.) It is, indeed, constant along the 
symplectic leaves of $\pi$, as the fields $w_{\pm}$ are not present in this new Lagrangian at all.
The new BV action is:
\begin{align}\label{MinimalisticMasterAction}
\widetilde{S}_{\rm BV}\;=\; 
\int\;d\tau\;d\sigma \; \mbox{Str}\Big( & \;
      {1\over 2}J_{2+}J_{2-} + {3\over 4}J_{1+}J_{3-} + {1\over 4}J_{3+}J_{1-} - J_{1+}{\bf P}_{31}J_{3-} \;+\;
\nonumber\\   
&
      + \sum\limits_{Z\in\{x,\theta,\lambda\}} (QZ)Z^{\star}  
   \Big)
\end{align}
where $Z^i$ runs over $\theta,x,\lambda$ and the action of $Q$ on $Z^i$ is the same as it was in the
original $\sigma$-model. 
The new BV phase space is smaller, it only contains $\theta,x,\lambda,\theta^{\star},x^{\star},\lambda^{\star}$. The BRST operator is 
now nilpotent off-shell; the dependence of the BV action on the antifields is linear.
The fields $\lambda_{L|R}$ enter only through their combination invariant under local rescalings
(they enter through ${\bf P}_{31}$).
This in particular implies that the BRST symmetry $Q$ is now a {\em local} symmetry.

Of course, the new action (\ref{NewAction}) is degenerate.

\section{Action of diffeomorphisms}\label{sec:diffeomorphisms}
\subsection{Formulation of the problem}\label{sec:FormulationOfTheProblem}
Let $L_2^{\star}$ be the BV Hamiltonian generating the left shift by elements of ${\bf g}_{\bar{2}}$; if $f$ is any
function of $g$, then:
\begin{equation}\label{DefLStar}
\left\{\mbox{Str}(A_2L_2^{\star})\,,\, f\right\}_{\rm BV} (g)\;=\; 
\left.{d\over dt}\right|_{t=0}f\left(e^{tA_2} g\right)
\end{equation}
The $L_0^{\star}$, $L^{\star}_1$ and $L^{\star}_3$ are defined similarly. In particular:
\begin{align}\label{BRSTOperator}
& \{\widetilde{S}_{\rm BV},\_\} \;=\;  \int \mbox{Str}\left(\lambda_3 L^{\star}_1 + \lambda_1 L^{\star}_3\right)
\end{align}
With these notations, when $X$ and $Y$ are two even elements of ${\bf g}$,
\begin{align}
\left\{\int \mbox{Str}(XL^{\star}),\int\mbox{Str}(YL^{\star}) \right\}_{\rm BV}\;=\;&
- \int\mbox{Str}([X,Y] L^{\star})
\end{align}
(Even elements are generators of ${\bf g}_2$ and ${\bf g}_0$, and also the generators of ${\bf g}_3$ and ${\bf g}_1$ multiplied 
by a Grassmann odd parameter.)

The infinitesimal action of diffeomorphisms is generated by the following BV Hamiltonian $V_{\xi}$:
\begin{align}\label{VXi}
&V_{\xi}  =\int \mbox{Str}\Big(  
(\iota_{\xi}D_0\lambda_3)\lambda_1^{\star} + (\iota_{\xi}D_0\lambda_1)\lambda_3^{\star} - 
(\iota_{\xi}J_3)L_1^{\star} - (\iota_{\xi}J_1)L_3^{\star} - (\iota_{\xi}J_2) L_2^{\star}
\Big)
\\   
& \mbox{\tt\small where } D_0\lambda = d\lambda + [J_0,\lambda]
\end{align}
In this section we will construct $\Phi_{\xi}$ such that:
\begin{equation}
V_{\xi} = \{\widetilde{S}_{\rm BV},\Phi_{\xi}\}_{\rm BV}
\end{equation}
It is very easy to construct such $\Phi_{\xi}$ if we don't care about the global symmetries
of $AdS_5\times S^5$. (Something like $\Phi_{\xi} = {\theta^{\alpha}\over\lambda^{\alpha}}V_{\xi}$.) But we will construct a $\Phi_{\xi}$ invariant
under the supersymmetries  of $AdS_5\times S^5$, {\it i.e.} invariant under the right shifts of $g$.
We believe that such an invariant construction has better chance of satisfying the equivariance 
conditions of \cite{Mikhailov:2016myt,Mikhailov:2016rkp} at the quantum level, because supersymmetries restrict quantum corrections. 
In particular, the equivariance condition must require that the $\Phi_{\xi}$ correspond, in some sense,
to a primary operator.

\paragraph     {Comment on gauge transformations}
In this Section we discuss vector fields on the factorspace $PS\;AdS$ defined by Eq. (\ref{PSAdS}).
They are the same as $SO(1,4)\times SO(5)$-invariant vector fields on
$C_L\times C_R\times PSU(2,2|4)$ modulo $SO(1,4)\times SO(5)$-invariant vertical vector fields.
All the formulas here are modulo vertical $SO(1,4)\times SO(5)$-invariant vector fields.

\subsection{Subspaces associated to a pair of pure spinors}
\underline{We use the notations of Section \ref{sec:Subspaces}}. For  $X_3\in [{\bf g}_{2L},\lambda_1]$ and $X_1\in [{\bf g}_{2R},\lambda_3]$, 
let ${\bf T}_2(X_1+X_2)$ denote the map:\marginpar{${\bf T}_2$}
\begin{align}
{\bf T}_2\;:\;&[\lambda_1,{\bf g}_{2L}]\oplus 
[\lambda_3,{\bf g}_{2R}]\longrightarrow {\bf g}_{2L}\oplus {\bf g}_{2R}
\\[5pt]   
{\bf T}_2&([\lambda_1,v_{2L}] + [\lambda_3,v_{2R}]) = v_{2L} + v_{2R}
\end{align}
\marginpar{{\tiny notations}\\ {\small Sec \ref{sec:Projector}}}(This is a direct sum of two completely independent linear maps.) 
For a pair $I_3\oplus I_1\in T^{\perp}C_R \oplus T^{\perp}C_L$ we decompose\footnote{For example, $I_3^{\rm split}$ denotes the component of $I_3$ which belongs to $[\lambda_1,{\bf g}_{2L}]$; the label ``split'' is because we could not invent any better notation.}:
\begin{align}
I_3\oplus I_1 \;=\;& 
(I_3^{\rm split}\oplus I_1^{\rm split}) + (I_3^{\rm ker}\oplus I_1^{\rm ker}) + (I_3^{\rm coker}\oplus I_1^{\rm coker})
\\    
\mbox{\tt\small where }&
I_3^{\rm split}\oplus I_1^{\rm split} \in [\lambda_1,{\bf g}_{2L}] \oplus [\lambda_3,{\bf g}_{2R}]
\\    
& 
I_3^{\rm ker}\oplus I_1^{\rm ker}
\in \mbox{ker}\;\Big[
T^{\perp}C_R \oplus T^{\perp}C_L 
\stackrel{(+)\circ(\{\lambda_3,\_\}\oplus\{\lambda_1,\_\})}{\longrightarrow}
{\bf g}_2
\Big]
\\   
&
I_3^{\rm coker}\oplus I_1^{\rm coker}
\in \mbox{coker}\;\Big[
{\bf g}_2
\stackrel{\{\lambda_3,\_\}+\{\lambda_1,\_\}}{\longrightarrow}
T^{\perp}C_R \oplus T^{\perp}C_L 
\Big]
\end{align}
where we must use a special representative of the  cokernel:
\begin{align}
I_3^{\rm ker} \;=\;&{1\over 2}
{\mbox{Tr}(\lambda_3 I_3 - \lambda_1 I_1)\over
\mbox{Tr}\left([\lambda_1,\lambda_3]_{\stl}\right)^2}
\left[\lambda_1,[\lambda_3,\lambda_1]_{\stl}\right]
\\   
I_1^{\rm ker} \;=\;&{1\over 2}
{\mbox{Tr}(\lambda_3 I_3 - \lambda_1 I_1)\over
\mbox{Tr}\left([\lambda_1,\lambda_3]_{\stl}\right)^2}
\left[\lambda_3,[\lambda_3,\lambda_1]_{\stl}\right]
\\  
I_3^{\rm coker} \;=\;&{1\over 2}
{\mbox{Tr}(\lambda_3 I_3 + \lambda_1 I_1)\over
\mbox{Tr}\left([\lambda_1,\lambda_3]_{\stl}\right)^2}
\left[\lambda_1,[\lambda_3,\lambda_1]_{\stl}\right]
\\   
I_1^{\rm coker} \;=\;&-{1\over 2}
{\mbox{Tr}(\lambda_3 I_3 + \lambda_1 I_1)\over
\mbox{Tr}\left([\lambda_1,\lambda_3]_{\stl}\right)^2}
\left[\lambda_3,[\lambda_3,\lambda_1]_{\stl}\right]
\end{align}
Similarly, any $I_2\in {\bf g}_2$ (assumed to be both TL and STL) can be decomposed:
\begin{align}
I_2 \;=\; & I_2^{\rm L} + I_2^{\rm R} + I_2^{\rm ker} + I_2^{\rm coker}
\\    
\mbox{\tt\small where } & 
I_2^{\rm L}\;\in\; {\bf g}_{2L} 
\\   
& I_2^{\rm R}\;\in\; {\bf g}_{2R}
\\   
& I_2^{\rm ker}\;\in\; \mbox{ker}\left[
  {\bf g}_2
  \stackrel{\{\lambda_3,\_\} + \{\lambda_1,\_\}}{\longrightarrow}
  T^{\perp}C_L \oplus T^{\perp}C_R
\right]
\\  
& I_2^{\rm coker}\;\in\; \mbox{coker}\left[
  T^{\perp}C_R \oplus T^{\perp}C_L
  \stackrel{(+)\circ(\{\lambda_3,\_\}\oplus\{\lambda_1,\_\})}{\longrightarrow}
  {\bf g}_2
\right]
\end{align}
Explicitly:
\begin{align}
I_2^{\rm ker}\;=\; &
{\mbox{Tr}(I_2\overline{[\lambda_3,\lambda_1]})\over\mbox{Tr} ([\lambda_3,\lambda_1]_{\stl})^2}
\overline{[\lambda_3,\lambda_1]}_{\rm TL}
\\    
I_2^{\rm coker}\;=\; &
{\mbox{Tr}(I_2[\lambda_3,\lambda_1])\over\mbox{Tr} ([\lambda_3,\lambda_1]_{\stl})^2}
[\lambda_3,\lambda_1]_{\rm STL}
\end{align}

\subsection{Construction of $\Phi_{\xi}$}
The generating function $V_{\xi}$ of the infinitesimal worldsheet diffeomorphisms (= vector fields) 
$\xi = \xi^{\tau}\partial_{\tau} + \xi^{\sigma}\partial_{\sigma}$, given by Eq. (\ref{VXi}),  is BV-exact:
\begin{align}
V_{\xi}  \;=\; \{\widetilde{S}_{\rm BV}, \Phi_{\xi}\} &
\label{VXiAgain}\\     
\Phi_{\xi} \;=\; - \int \mbox{Str}\Big( &  ({\bf P}_{31} \iota_{\xi}J_3)\lambda_1^{\star} + ({\bf P}_{13} \iota_{\xi}J_1)\lambda_3^{\star}\;+
\nonumber\\   
& + \Big({\bf T}_2\left(
   \iota_{\xi}J_3^{\rm split} + 
   \iota_{\xi}J_1^{\rm split} 
\right)  + {\cal A}[\lambda_3,\lambda_1]_{\stl} \Big)L_2^{\star}\;+
\nonumber\\   
& + {\cal B}\left(
[\lambda_1,[\lambda_3,\lambda_1]_{\stl}]L_3^{\star} - [\lambda_3,[\lambda_3,\lambda_1]_{\stl}]L_1^{\star}
\right)
\Big)
\label{MinimalPhi}
\\
& \mbox{\tt\small where }
\nonumber \\  
& \iota_{\xi} J \;=\; - \xi^{\alpha}\partial_{\alpha}g g^{-1}
\nonumber\\  
& {\cal A}\;=\; {1\over 2}{\mbox{Tr}\left(
\lambda_3(1-{\bf P}_{31})\iota_{\xi}J_3 - \lambda_1 (1-{\bf P}_{13})\iota_{\xi}J_1
\right) \over \mbox{Tr}([\lambda_3,\lambda_1]_{\stl})^2}
\nonumber \\  
& {\cal B}\;=\;
{\mbox{STr}([\lambda_3,\lambda_1]\iota_{\xi}J_2)
\over \mbox{STr}(\lambda_3\lambda_1)\;\mbox{Tr}([\lambda_3,\lambda_1]_{\stl})^2}
\end{align}
The coefficients $\cal A$ and $\cal B$ satisfy:
\begin{align}
& (Q{\cal A})[\lambda_3,\lambda_1]_{\stl} \;=\; \iota_{\xi}J_2^{\rm coker}
\\    
& {\cal A} \left(
  [\lambda_1,[\lambda_3,\lambda_1]_{\stl}] + [\lambda_3,[\lambda_3,\lambda_1]_{\stl}]
\right)
\;=\;\iota_{\xi} J_3^{\rm ker} + \iota_{\xi}J_1^{\rm ker}
\\   
& (Q{\cal B})\left(
[\lambda_1,[\lambda_3,\lambda_1]_{\stl}] - [\lambda_3,[\lambda_3,\lambda_1]_{\stl}]
\right)\;=\; \iota_{\xi}J_1^{\rm coker} + \iota_{\xi}J_3^{\rm coker}
\\    
& {\cal B}\left(
  \{\lambda_3,[\lambda_1,[\lambda_3,\lambda_1]_{\stl}]\} -
  \{\lambda_1,[\lambda_3,[\lambda_3,\lambda_1]_{\stl}]\}
\right)\;=\;\iota_{\xi}J_2^{\rm ker}
\end{align}
Eq. (\ref{VXiAgain}) follows from:
\begin{align}
\iota_{\xi}J \;=\; &
{\bf P}_{31}\iota_{\xi}J_3 + {\bf P}_{13}\iota_{\xi}J_1 +
\nonumber\\   
& + \iota_{\xi}J_3^{\rm split} + 
    \iota_{\xi}J_1^{\rm split} +
    \iota_{\xi}J_3^{\rm ker} +
    \iota_{\xi}J_1^{\rm ker} + 
\nonumber\\     
& + (Q{\cal B}) \left(
[\lambda_1,[\lambda_3,\lambda_1]_{\stl}] - [\lambda_3,[\lambda_3,\lambda_1]_{\stl}]
\right) + 
\nonumber\\  
& + \iota_{\xi}J_2^{\rm split} + \iota_{\xi}J_2^{\rm ker} +  (Q{\cal A})[\lambda_3,\lambda_1]_{\stl}
\end{align}

\paragraph     {Some useful identities}
\begin{align}
\mbox{STr}\left([\lambda_3,[\lambda_3,\lambda_1]_{\stl}]\;(1-{\bf P}_{31})\iota_{\xi}J_3\right)\;=\;&
-{\mbox{STr}(\lambda_3\lambda_1)\over 2}\mbox{Tr}\left(\lambda_3(1-{\bf P}_{31})\iota_{\xi}J_3\right)
\end{align}
\begin{align}
& \{\lambda_1,[\lambda_3,[\lambda_3,\lambda_1]_{\stl}]\} = 
- \{\lambda_3,[\lambda_1,[\lambda_3,\lambda_1]_{\stl}]\} =
\nonumber\\   
\;=\; &
{1\over 2}\overline{[\lambda_1,\lambda_3]}_{\tl}\mbox{Str}(\lambda_3\lambda_1)  + 
{1\over 8}\left(
        (\mbox{Str}(\lambda_3\lambda_1))^2 
        - 2\mbox{Tr}[\lambda_1,\lambda_3]^2
\right) {\bf 1}\;=\;
\nonumber\\   
\;=\; & 
{1\over 2}\overline{[\lambda_1,\lambda_3]}_{\tl}\mbox{Str}(\lambda_3\lambda_1)  -
{1\over 4}\mbox{Tr}([\lambda_3,\lambda_1]_{\stl})^2\;{\bf 1}
\end{align}
\begin{align}
\mbox{STr}
& \Big([\lambda_1,[\lambda_3,\lambda_1]_{\stl}]\;[\lambda_3,[\lambda_3,\lambda_1]_{\stl}]\Big)=
\nonumber\\   
\;=\; &
-{1\over 2}\mbox{Str}(\lambda_3\lambda_1)\mbox{Str}\Big([\lambda_1,\lambda_3]_{\stl}\overline{[\lambda_1,\lambda_3]}_{\tl}\Big)\;=\;
\nonumber\\   
\;=\; &
-{1\over 2}\mbox{Str}(\lambda_3\lambda_1)\mbox{Tr}([\lambda_3,\lambda_1]_{\stl})^2
\end{align}
Notice that we have $\mbox{Tr}([\lambda_3,\lambda_1]_{\stl})^2$ in denominators. At the same time:
\begin{equation}
\mbox{STr}([\lambda_3,\lambda_1]_{\stl})^2 = \mbox{STr}([\lambda_3,\lambda_1])^2 = 0
\end{equation}

\section{Regularization}\label{sec:Regularization}
The ``minimalistic action'' (\ref{MinimalisticMasterAction}) cannot be regularized in a way that would preserve the
symmetries of $AdS_5\times S^5$;  it is impossible to choose a $PSU(2,2|4)$-invariant 
Lagrangian submanifold so that the restriction of the Master Action of Eq. (\ref{MinimalisticMasterAction}) to it be
non-degenerate. Let us therefore return to the original action of Eqs. (\ref{ActionSplit}), (\ref{StandardBV}), but in a way
preserving the worldsheet diffeomorphisms. The construction is somewhat similar to the
description of the topological A-model in \cite{Alexandrov:1995kv}.

\subsection{Adding more fields}\label{sec:AddingOmega}
Add a pair of bosonic 1-form fields $\omega_3$ and $\omega_1$, 
taking values in ${\bf g}_3$ and ${\bf g}_1$, respectively, and their antifields $\omega_1^{\star}$ and $\omega_3^{\star}$, also 1-forms:
\begin{equation}
\omega_{\rm BV} = \int \mbox{STr}\left( d\omega^{\star}_3\wedge d\omega_1 + d\omega^{\star}_1\wedge d\omega_3\right)
\end{equation}
(where $d$ is the differential in the field space, not on the worldsheet!). In other words,
for any ``test 1-forms'' $f_1$ and $g_3$:
\begin{align}
  \left\{\int\mbox{Str}\;f_1\wedge \omega_3^{\star}\;,\;\int\mbox{Str}\;\omega_1\wedge g_3\right\}
  \;=\;
  \int\mbox{Str}\;f_1\wedge g_3
  \\
  \left\{\int\mbox{Str}\;g_3\wedge \omega_1\;,\;\int\mbox{Str}\;\omega_3^{\star}\wedge f_1\right\}
  \;=\;
  \int\mbox{Str}\;g_3\wedge f_1
\end{align}
We define the BV Master Action as follows:\marginpar{$\widetilde{S}^{+}_{\rm BV}$}
\begin{align}\label{HatSBV}
& \widetilde{S}^{+}_{\rm BV} = \widetilde{S}_{\rm BV} + \int \mbox{STr}(\omega_3^{\star} \wedge\omega_1^{\star})
\end{align}
and the BV Hamiltonian for the action of diffeomorphisms as follows:
\begin{align}
\widehat{V}_{\xi} \;=\;& \{\widetilde{S}^{+}_{\rm BV}\,,\,\widehat{\Phi}_{\xi}\}_{\rm BV}
\\    
\widehat{\Phi}_{\xi} \;=\; & \Phi_{\xi} + \int \mbox{STr}(\omega_3\wedge {\cal L}_{\xi}\omega_1)
\label{TotalPhi}
\end{align}
where ${\cal L}_{\xi}$ is the Lie derivative.

\vspace{10pt}
\noindent
The expression $\int \mbox{STr}(\delta\omega_3\wedge \delta\omega_1)$ defines a symplectic structure on the space of 1-forms 
with values in ${\bf g}_{\rm odd}$. The expression $\int \mbox{STr}(\omega_3^{\star} \wedge\omega_1^{\star})$ is the 
\amklink{math/bv/omega/Duistermaat-Heckman_formula.html}{corresponding Poisson bivector}.
The Lie derivative preserves this (even) symplectic structure, and $\int \mbox{STr}(\omega_3\wedge {\cal L}_{\xi}\omega_1)$ is 
the corresponding Hamiltonian.

\subsection{A canonical transformation}
Let us do the canonical transformation by a flux of the following odd Hamiltonian:
\begin{equation}\label{PsiA}
   \Psi_{(0)} \;=\;  \int  \mbox{STr}\;  [A_0, \omega_3] \wedge\omega_1 \;=\;
   - \int \mbox{STr}\; [A_0,\omega_1]\wedge\omega_3
\end{equation}
This is the Hamiltonian of $[A_0,\_]$ in the same sense as $\int \mbox{STr}(\omega_3\wedge {\cal L}_{\xi}\omega_1)$ is the 
Hamiltonian of ${\cal L}_{\xi}$; we again use the same procedure of passing from Eq. (\ref{OriginalAction}) to Eq. (\ref{NewDarbouxCoordinates}), 
actually in reverse. 

The effect of the flux of $\Psi_{(0)}$ on the BV Master Action $\widetilde{S}^+_{\rm BV}$ of Eq. (\ref{HatSBV}) is:
\begin{align}
  \widetilde{S}^{+}_{\rm BV}\;=\;
  & \widetilde{S}_{\rm BV} + 
    \int \mbox{STr}\;\omega_3^{\star}\wedge \omega_1^{\star}
    \nonumber\\[5pt]    
  \mbox{\tt\small becomes}
  &  \nonumber\\[5pt]   
  \widetilde{S}'_{\rm BV}\;=\;
  & \widetilde{S}_{\rm BV}  + 
    \int \mbox{STr}\;
    \left[\left(\{\lambda_3,\lambda_1\} - {1\over 2}[A_0,A_0]\right),\;\omega_3\right]
    \wedge
    \omega_1\;+
    \nonumber\\    
  & + \int \mbox{STr}
    (\omega_3^{\star} + [A_0,\omega_3])\wedge (\omega_1^{\star} - [A_0,\omega_1])
    \;=\;
    \nonumber\\
  \;=\;
  & \widetilde{S}^+_{\rm BV}  +
    \int \mbox{STr}\;\{\lambda_3,\lambda_1\}
    \{\omega_{3},\wedge\omega_{1}\} \;+
    \label{NewBVActionSimplified}\\   
  & + \int \mbox{STr}\Big(
    [A_0,\omega_3]\wedge\omega_1^{\star} +
    [A_0,\omega_1]\wedge\omega_3^{\star}
    \Big) 
    \nonumber
\end{align}
Notice that the terms of the form $A^2\omega^2$ cancelled. This is automatic, because such terms 
would contradict the Master Equation (the bracket $\left\{\omega^{\star}\omega^{\star},A^2\omega^2\right\}$ would have nothing to 
cancel against). 

The purpose of this canonical transformation was, essentially, to introduce the compensator
term $[A,\omega]$ into the action of $Q$ on $\omega$, {\it cp.} Eq. (\ref{QgWithA0}). We will discuss this in a more
general context in Section \ref{sec:GluingCharts}.

We are now ready to construct the Lagrangian submanifold.

\subsection{Constraint surface and its conormal bundle}\label{sec:ConormalBundle}
The configuration space $X$ of this new theory is parametrized by 
$g,\lambda_3$,$\lambda_1$,$\omega_{3\pm}$ and $\omega_{1\pm}$. Let us consider a subspace $Y\subset X$ defined by the constraints:
\begin{align}
(1-{\bf P}_{13})\omega_{1+} \;=\; & 0
\nonumber\\    
(1-{\bf P}_{31})\omega_{3-} \;=\; & 0
\label{ConstraintSurface}\\   
\omega_{1-} \;=\; & 0
\nonumber\\     
\omega_{3+} \;=\; & 0
\nonumber
\end{align}
Consider the 
\amklink{math/bv/BRST-formalism/Family_of_Lagrangian_submanifolds.html\#(part.\_.Conormal\_bundle)}{odd conormal bundle}
$\Pi T^{\perp}Y$ of $Y\subset X$ in the BV phase space $\Pi T^*X$. 
As any conormal bundle, this is a Lagrangian submanifold. \marginpar{$\Pi T^{\perp}Y$}
The restriction of $\widetilde{S}'_{\rm BV}$ on this Lagrangian submanifold is still degenerate. But let us deform 
it by the following  generating function:
\begin{equation}\label{GaugeFermion}
\Psi = \int  \mbox{STr}\Big(
   \omega_{3-}{\bf P}_{13}J_{1+} + \omega_{1+}{\bf P}_{31}J_{3-}
\Big)
\end{equation}
The restriction of $\widetilde{S}'_{\rm BV}$ to this deformed Lagrangian submanifold is equal to:
\begin{align}
\int\mbox{STr}\Big(&
{1\over 2}J_{2+}J_{2-} + {3\over 4}J_{1+}J_{3-} + {1\over 4}J_{3+}J_{1-} 
\;+
\nonumber\\   
& +  w_{1+}D_{0-}\lambda_3 + w_{3-}D_{0+}\lambda_1 + N_{0+}N_{0-} \;+ 
\omega_{3+}^{\star} \omega_{1-}^{\star} + 
\omega_{3-}^{\star} \omega_{1+}^{\star}
\Big)
\end{align}
where $N_{0+} = \{w_{1+},\lambda_3\}$, $N_{0-} = \{w_{3-},\lambda_1\}$,
\begin{equation}
w_{1+} = {\bf P}_{13}\omega_{1+} \mbox{ \tt\small and } w_{3-} = {\bf P}_{31}\omega_{3-}
\end{equation}
Notice that the terms:
\begin{align}
\int \mbox{STr}\Big( 
& 
[A_0,\lambda_3]\lambda_3^{\star} + [A_0,\lambda_1]\lambda_1^{\star} \;+ 
\nonumber\\   
+ & 
\;[A_0,\omega_{3+}]\omega_{1-}^{\star} + [A_0,\omega_{3-}]\omega_{1+}^{\star} +
[A_0,\omega_{1+}]\omega_{3-}^{\star} + [A_0,\omega_{1-}]\omega_{3+}^{\star} 
\Big)
\end{align}
vanish on $T^{\perp}Y$. Indeed, the vector field:
\begin{align}
& [A_0,\lambda_3]{\partial\over\partial\lambda_3} + [A_0,\lambda_1]{\partial\over\partial\lambda_1} \;+
\nonumber\\   
+ \;
& [A_0,\omega_{3+}]{\partial\over\partial\omega_{3+}} 
+ [A_0,\omega_{3-}]{\partial\over\partial\omega_{3-}}
+ [A_0,\omega_{1+}]{\partial\over\partial\omega_{1+}} 
+ [A_0,\omega_{1-}]{\partial\over\partial\omega_{1-}}
\end{align}
is tangent to the constraint surface (\ref{ConstraintSurface}); the conormal bundle, by definition, consists of 
those one-forms which vanish on such vectors.
The term  $\omega_{3+}^{\star} \omega_{1-}^{\star} + \omega_{3-}^{\star} \omega_{1+}^{\star}$ computes the contrubution to the action from the fiber
$\Pi T^{\perp}Y$. The coordinates of the fiber enter without derivatives, and decouple.

\vspace{10pt}
\noindent
We therefore return to the original $AdS_5\times S^5$ action of Eq. (\ref{ActionSplit}).

\vspace{10pt}
\noindent
But now we understand how the worldsheet diffeomorphisms act, at the level of the BV phase space.

\subsection{Gluing charts}\label{sec:GluingCharts}
In our construction we used a lift of  $AdS_5\times S^5$ to $PSU(2,2|4)$ (Section \ref{sec:A0}). 
This is only possible locally. Therefore, we have to explain how to glue together overlapping 
patches. This is a particular case of a general construction, which we will now describe.

\paragraph     {The idea} is to build a theory which is {\em locally} (on every patch of $AdS_5\times S^5$)
a direct product of two theories $S_{(\phi)}$ and $S_{(w)}$:
\begin{equation}
S_{\rm tot} \;=\; 
S_{(\phi)} + S_{(w)} = S_{\rm cl}(\phi) + Q^{\mu}(\phi)\phi^{\star}_{\mu} \;\; + \;\;
{1\over 2}w^{\star}_a (\omega^{-1})^{ab} w^{\star}_b
\end{equation}
but transition functions between overlapping patches mix $\phi$ and $w$.

\vspace{10pt}
\noindent
Consider the following  data, consisting of two parts. The first part is a Lie group $H$
and a principal $H$-bundle $E$ with base $B$. Suppose that $B$ comes with a nilpotent vector
field $Q\in \mbox{Vect}(B)$ and a 
$Q$-invariant action $S_{\rm cl}\in \mbox{Fun}(B)$. Then $S_B(\phi,\phi^{\star}) = S_{\rm cl}(\phi) + Q^{\mu}(\phi)\phi_{\mu}^{\star}$ satisfies 
the Master Equation on the BV phase space $\Pi T^*B$. 
The second part of the data is a symplectic vector space $W$ which is a representation of $H$. 
This means that $W$ is equipped with an {\em even}  $H$-invariant symplectic form $\omega$.

Let us cover $B$ with charts $\{U_{i}| i\in {\cal I}\}$ and trivialize $E$ over each chart:
\begin{equation}
p^{-1}(U_{i}) \simeq U_{i}\times H
\end{equation}
At the intersection $U_{i}\cap U_{j}$ we identify $(\phi,h_{i})\in U_{i}\times H$ with 
$(\phi,h_{j})\in U_{j}\times H$ if
\begin{equation}
h_{j} = u_{ji}(\phi)h_{i}
\end{equation}
All this comes from $E\stackrel{H}\rightarrow B$. We will now construct a new odd symplectic manifold, which is
locally $\Pi T^*U_{j}\times \Pi T^*W$, with some  transition functions, which we will now describe. 

\paragraph     {Technical assumption:} in this Section we assume that all $w$ are bosons, and
that $H$ is a ``classical'' ({\it i.e.} not super) Lie group. This is enough for our
considerations.

\paragraph     {Transition functions} Let $\bf h$ be the Lie algebra of $H$. For each $\alpha\in \mbox{Map}(B,{\bf h})$
consider the following BV Hamiltonian:
\begin{align}
\chi_{\alpha} \;=\; & \left\{S_{\rm tot} , F_{\alpha}\right\}
\label{DefHXi}\\    
\mbox{\tt\small where } F_{\alpha} \;=\; & - {1\over 2}w^b\rho_*(\alpha(\phi))_b^a\;\omega_{ac}\;w^c
\label{DefFXi}
\end{align}
Here $\rho_*$ is the representation of the Lie algebra corresponding to the representation $\rho$ of the
group, and $\omega$ is the symplectic form of $W$. Eq. (\ref{DefFXi}) defines $F_{\alpha}$ as the
Hamiltonian of the infinitesimal action of $\alpha$ on $w$, {\it i.e.} the ``usual'' (even) moment map. 
(Here we use our assumption that $\omega$ is $H$-invariant.) The explicit formula for $\chi_{\alpha}$ is:
\begin{equation}
\chi_{\alpha} = \rho_*(\alpha(\phi))^a_bw^b \;w^{\star}_a
\;-\; {1\over 2} w^b\rho_*(Q\alpha(\phi))^a_b\omega_{ac}w^c
\end{equation}
Notice that:
\begin{equation}\label{EquivarianceOfF}
\left\{\chi_{\alpha_1}\;,\;F_{\alpha_2}\right\}\;=\;- F_{[\alpha_1,\alpha_2]}
\end{equation}
The flux of the BV-Hamiltonian vector field $\{\chi_{\alpha},\_\}$ is a canonical transformation, 
and Eq. (\ref{DefHXi}) implies that this canonical transformation is a symmetry of $S_{\rm tot}$. This canonical 
transformation does not touch $\phi^{\mu}$, it only acts on $\phi^{\star}, w, w^{\star}$. We identify 
$(\phi, \phi^{\star}_{i}, w_{i}, w^{\star}_{i})$ on chart $U_{(i)}$ with $(\phi,\phi^{\star}_{j}, w_{j}, w^{\star}_{j})$ on chart $U_{(j)}$ when 
$(\phi^{\star}_{j}, w_{j}, w^{\star}_{j})$ is the flux of  $(\phi_i^{\star}, w_{i}, w^{\star}_{i})$ by the time $1$ along the vector field 
$\{\chi_{\alpha_{ji}},\_\}$ where $\alpha_{ji}$ is the log of $u_{ji}$, {\it i.e.} $u_{ji} = e^{\alpha_{ji}}$. Explicitly:
\begin{align}
w_{j}^a \;=\; & \rho\left(u_{ji}\right)^a_bw^b_{i}
\label{RotationOfW}\\    
w^{\star}_{ja} \;=\; & 
\rho\left(u_{ji}^{-1}\right)_a^b\; w^{\star}_{ib} 
\; - \; \omega_{ab}\;Q\rho\left(u_{ji}\right)^b_c \; w^c_{i}
\label{RotationAndShiftOfWStar}\\    
\phi^{\star}_{j\mu} \;=\; & \phi^{\star}_{i\mu} -
w^{\star}_{ja}\rho_*\left(u_{ji}\stackrel{\leftarrow}{\partial \over\partial\phi^{\mu}}u_{ji}^{-1}\right)_b^aw_{j}^b 
- {1\over 2}w^a_{j}\omega_{ab} 
{\partial\over\partial\phi^{\mu}}\rho_*\left(Qu_{ji} u_{ji}^{-1}\right)^b_cw_{j}^c
\label{ShiftOfPhiStar}
\end{align}
These gluing rules are consistent on triple intersections because of Eq. (\ref{EquivarianceOfF}).

\paragraph     {Lagrangian submanifold}
Eqs. (\ref{RotationAndShiftOfWStar}) and (\ref{ShiftOfPhiStar}) look somewhat unusual. In particular, the ``standard'' 
Lagrangian submanifold\footnote{In BV formalism, there is no such thing as the standard 
Lagrangian submanifold. We invented this notion to denote the one where all antifields
(w.r.to some Darboux coordinates) are zero. This is often a useful starting point to construct
Lagrangian submanifolds.}
$\phi^{\star}=w^{\star}=0$ is not well-defined, because it is incompatible with our transition functions.
One simple example of a well-defined Lagrangian submanifold is $w=\phi^{\star}=0$. We will now 
give another example, which repairs the ill-defined $w^{\star}=\phi^{\star}=0$.

The construction requires a choice of a {\em connection}  in the principal bundle $E\stackrel{H}\rightarrow B$.
To specify a connection, we choose on every chart $U_{i}$ some ${\bf h}$-valued 1-form $A_{i\mu}$,
with the following identifications on the intersection $U_i\cap U_j$:
\begin{equation}\label{GluingOfConnection}
{\partial\over\partial \phi^{\mu}} + A_{j\mu}(\phi) = 
u_{ji}(\phi)\left({\partial\over\partial \phi^{\mu}} + 
A_{i\mu}(\phi)\right)(u_{ji}(\phi))^{-1}
\end{equation}
and in particular:
\begin{equation}
Q\rho(u_{ji})^a_b\;+\; 
\rho_*(Q^{\mu}A_{j\mu})^a_c\rho\left(u_{ji}\right)^c_b \;-\; 
\rho(u_{ji})^a_c\rho_*(Q^{\mu}A_{i\mu})^c_b \;=\;0
\end{equation}
On every chart, let us pass to a new set of Darboux coordinates, by doing the canonical
transformation with the following gauge fermion:
\begin{equation}
\Psi_{i} =  {1\over 2}w^a_{i}\;\omega_{ab} \; Q^{\mu}(\phi)\rho_*(A_{i\mu}(\phi))^b_c\; w^c_{i}
\end{equation}
Notice that $\Psi_{i}$ does not depend on antifields; therefore this canonical transformation
is just a shift:
\begin{align}
\tilde{w}_i^{a}\;=\;& w_i^{a}
\label{wtilde}\\    
\tilde{w}^{\star}_{ia}\;=\;&w^{\star}_{ia} \;-\;
\omega_{ab} \; Q^{\nu}(\phi)\rho_*(A_{i\nu}(\phi))^b_c\; w^c_{i}
\label{wstartilde}\\    
\tilde{\phi}^{\star}_{i\mu} \;=\;&\phi^{\star}_{i\mu} \;-\;
{1\over 2}{\partial\over\partial\phi^{\mu}}\Big[
  w^a_{i}\;\omega_{ab} \; Q^{\nu}(\phi)\rho_*(A_{i\nu}(\phi))^b_c\; w^c_{i}
                                   \Big]
\label{phistartilde}
\end{align}
This canonical transformation does not preserve $S_{\rm BV}$, therefore
the {\em expression} for the action will be different in different charts, see Eq. (\ref{NewBVActionSimplified}). In
particular, it will contain the term $\tilde{w}^{\star}Q^{\mu}\rho_*(A_{i\mu})\tilde{w}$, which means that the action of the BRST
operator on $\tilde{w}$ involves the connection. On the other hand, the transition functions simplify:
\begin{align}
  \tilde{w}_{j}^a \;=\;
  & \rho\left(u_{ji}(\phi)\right)^a_b\tilde{w}^b_{i}
    \label{SimplifiedRotationOfW}\\    
  \tilde{w}^{\star}_{ja} \;=\;
  & \rho\left(u_{ji}(\phi)^{-1}\right)_a^b\; \tilde{w}^{\star}_{ib} 
    \label{SimplifiedRotationOfWStar}\\    
  \tilde{\phi}^{\star}_{j\mu} \;=\;
  & \tilde{\phi}^{\star}_{i\mu} - \tilde{w}^{\star}_{ic}
    \rho\left(u_{ji}(\phi)^{-1}\right)^c_a 
    \left(
    \rho\left(u_{ji}(\phi)
    \right)^a_b
    \stackrel{\leftarrow}{\partial\over\partial\phi^{\mu}}
    \right) \tilde{w}_{i}^b \;
    \label{SimplifiedPhiStar}
\end{align}
These are the usual transition functions of the odd cotangent bundle $\Pi T^* {\cal W}$, where $\cal W$
is the vector bundle with the fiber $W$, associated to the principal vector bundle  $E\stackrel{H}\rightarrow B$.

In particular, the ``standard'' Lagrangian submanifold $\tilde{w}^{\star}=\tilde{\phi}^{\star}=0$ is compatible with gluing.
The corresponding BRST operator is defined by the part of the BV action linear in the antifields:
\begin{equation}
Q_{\rm BRST} \;=\;Q^{\mu}{\partial\over\partial\phi^{\mu}} +  Q^{\nu}\rho_*(A_{\nu})^a_b\tilde{w}^b{\partial\over\partial \tilde{w}^a}
\end{equation}
After this canonical transformation of Eqs. (\ref{wtilde}), (\ref{wstartilde}) and (\ref{phistartilde}), the new $S_{\rm cl}$ is such that
this $Q_{\rm BRST}$ is nilpotent on-shell.

\paragraph     {Gluing together $\Phi_{\xi}$}
Let us consider the relation between the functions $\widehat{\Phi}_{\xi}$ defined by Eq. (\ref{TotalPhi}) on two overlapping
charts. It is enough to consider the case of infinitesimal transition function, {\it i.e.} 
$u_{ji} = {\bf 1} + \epsilon \alpha_{ji}$, where $\epsilon$ is infinitesimally small. With $F_{\alpha}$ defined in Eq. (\ref{DefFXi}),
the difference between $\widehat{\Phi}_{\xi}$ on two coordinate charts is:
\begin{align}
\delta_{ji} \widehat{\Phi}_{\xi} = \{\{S_{\rm tot},F_{\alpha_{ji}}\}, \widehat{\Phi}_{\xi}\} =
- \{\{S_{\rm tot},\widehat{\Phi}_{\xi}\},F_{\alpha_{ji}}\} + \{S_{\rm tot}, \{F_{\alpha_{ji}},\widehat{\Phi}_{\xi}\}\}
\end{align}
The first term on the RHS is zero:
\begin{equation}
\{\{S_{\rm tot},\widehat{\Phi}_{\xi}\},F_{\alpha_{ji}}\}=0
\end{equation}
since $F_{\alpha}$ is diffeomorphism-invariant. Let us study the second term. We have: 
\begin{align}
  & \{F_{\alpha_{ji}},\widehat{\Phi}_{\xi}\} = \{F_{\alpha_{ji}},\Phi_{\xi}\} = 
    {1\over 2}w^a\omega_{ab}\Phi_{\xi}^{\mu}{\partial\over\partial\phi^{\mu}}(\alpha_{ji})^b_cw^c\;=\;
    \nonumber\\   
  =\; & - {1\over 2}\left(w^a\omega_{ab}\Phi_{\xi}^{\mu}
        \left(
        \delta_{ji}A_{\mu}  -  [\alpha_{ji},A_{\mu}]\right)^b_cw^c
        \right)\;=\;
        \nonumber \\
  =\; & - F_{\Phi^{\mu}_{\xi} \delta_{ji}A_{\mu}} +
        {1\over 2}\;w^a\omega_{ab}\Phi_{\xi}^{\mu}[\alpha_{ji},A_{\mu}]^b_cw^c\;=
        \nonumber \\
  =\; & - F_{\Phi^{\mu}_{\xi} \delta_{ji}A_{\mu}} - \{\{S_{\rm tot},F_{\alpha_{ji}}\},F_{\Phi^{\mu}A_{\mu}}\}
\end{align}
where $A_{\mu}$ is any connection, transforming as in Eq. (\ref{GluingOfConnection}). Therefore the following expression:
\begin{equation}\label{PhiWithConnection}
\widehat{\Phi}'_{\xi} = \widehat{\Phi}_{\xi} + \{S_{\rm tot}, F_{\Phi_{\xi}^{\mu}A_{\mu}}\}
\end{equation}
is consistent on intersections of patches. 

The correcting term $\{S_{\rm tot}, F_{\Phi_{\xi}^{\mu}A_{\mu}}\}$ is the infinitesimal gauge transformation (see Eqs. (\ref{DefHXi})
and (\ref{DefFXi})) with the parameter $\Phi_{\xi}^{\mu}A_{\mu}$. 

\paragraph     {Back to $AdS_5\times S^5$}
In our case $B$ is the pure spinor bundle over super-$AdS_5\times S^5$; the coordinates $\phi$
are functions from the worldsheet to $PS\;AdS_5\times S^5$ (defined in Eq. (\ref{PSAdS})).
The total space $E$ is the space of maps from the worldsheet to $C_L\times C_R\times PSU(2,2|4)$.
Notice that $C_L\times C_R\times PSU(2,2|4)$ is a principal $H$-bundle over $PS\;AdS_5\times S^5$. 
It has a natural $PSU(2,2|4)$-invariant connection, which for every tangent vector:
\begin{equation}
(\dot{\lambda}_L,\dot{\lambda}_R,\dot{g})\in T\left(C_L\times C_R\times PSU(2,2|4)\right)
\end{equation}
declares its vertical component to be $(\dot{g}g^{-1})_{\bar{0}}$, {\it i.e.} the projection of $\dot{g}$ on the
denominator of (\ref{SuperAdS}) using the Killing metric. This defines, pointwise, the connection
on the space of maps.

\vspace{8pt}
\noindent
It is natural to use this connection as $A_{\mu}$ in Eq. (\ref{PhiWithConnection}).

\vspace{8pt}
\noindent
Notice that we do {\em not} need a connection to write the BV Master Action (Eq. (\ref{HatSBV})).
But the connection {\em is} needed to construct $\widehat{\Phi}_{\xi}'$ (and also in our construction of the
Lagrangian submanifold).

\section{Taking apart the AdS sigma model}\label{sec:TakingApart}
The standard action given by Eq. (\ref{ActionSplit}) depends on the worldsheet complex structure and is
polynomial in the pure spinor variables. In the BV formalism, it corresponds to a specific
choice of the Lagrangian submanifold. We can change the action to a physically equivalent one,
by adding BRST quartets and/or deforming the Lagrangian submanifold. We can ask ourselves,
what is the simplest formulation of the theory, in the BV language, preserving the
symmetries of $AdS_5\times S^5$? (Of course, the notion of ``being the simplest'' is somewhat
subjective.) In this paper we gave an example of such a ``minimalistic'' formulation:
\begin{align}
  S_{\rm BV} \;=\;
  & S_{(g, \lambda)} + S_{(\omega)} =
    \nonumber \\
  \;=\;
  & \int \mbox{STr}\left(
    J_1\wedge (1-{\bf P}_{31})J_3 - J_1\wedge {\bf P}_{31} J_3
    \right)
    \;+\;
    \nonumber \\  
  & + \int \mbox{STr}\left(
    \lambda_3 L^{\star}_1 + \lambda_1 L^{\star}_3
    \right)\;+
    \label{SeparatedBV} \\  
  & + \int {1\over 2}\mbox{STr}\left(\omega_3^{\star} \wedge  \omega_1^{\star}\right)
    \nonumber
\end{align}
Here $L^{\star}$ are the BV Hamiltonians of the left shift, Eq. (\ref{DefLStar}).
The relation of Eq. (\ref{SeparatedBV}) to the original BV action (\ref{StandardBV}) is through adding BRST quartet
(Section \ref{sec:Regularization}) and canonical transformations (Eqs. (\ref{GaugingAwayG}), (\ref{PsiA}), (\ref{GaugeFermion})). Subjectively, Eq. (\ref{SeparatedBV})
is the simplest way of presenting the worldsheet Master Action for $AdS_5\times S^5$.

The Master Action (\ref{SeparatedBV}) does not depend on the worldsheet metric.
The dependence on the worldsheet metric (through the complex structure) comes later when we
choose the Lagrangian submanifold.

The way Eq. (\ref{SeparatedBV}) is written, it seems that $w$ is completely
decoupled from $g$ and $\lambda$. But the transition functions on overlapping charts, described
in Section \ref{sec:GluingCharts}, do mix the two sets of fields.

The Master Action (\ref{SeparatedBV}) is non-polynomial in $\lambda$, because of ${\bf P}_{31}$.

\section{Generalization}\label{sec:Generalization}
Consider a sigma-model whose target space is some supermanifold $\cal X$.
Suppose that $\cal X$ is equipped with a nilpotent odd vector field $Q\in \mbox{Vect}({\cal X})$, generating 
a {\em gauge symmetry} of the sigma-model.
In minimalistic sigma-models {\em the BRST operator is just an odd nilpotent vector field 
on the target space}.

This means that the field configuration $X(\sigma,\tau)$ has the same action as $e^{\epsilon(\sigma,\tau)Q}X(\sigma,\tau)$ for 
an arbitrary odd gauge parameter function $\epsilon$ on the worldsheet:
\begin{equation}
S[X] = S[e^{\epsilon Q}X]
\end{equation}
Locally and away from the fixed points of $Q$ this implies that one of the target space fermionic 
coordinates completely decouples from the action (the action does not depend on it). 
In case of pure spinor sigma-model, this gauge symmetry does not account for {\em all} 
degeneracy of the action. All directions in the $\theta$ space tangent to the pure spinor cones
are degenerate directions of the quadratic part of the action. 

Let us add an additional scalar field on  
the worldsheet $\Lambda(\sigma,\tau)$ and consider the following solution of the Master Equation:
\begin{equation}
S_{\rm BV} = S + \int \Lambda Q^A(X) X^{\star}_A
\end{equation}
In the pure spinor case $\cal X$ is parametrized by $g\in PSU(2,2|4)$ and $\lambda_L$, $\lambda_R$ modulo 
rescaling ({\it i.e.} projective pure spinors). 

In Type II pure spinor theory, there are actually {\em two} anticommuting BRST symmetries, 
$Q_L$ and $Q_R$, and the term in $S_{\rm BV}$ linear in antifields is 
\begin{equation}
\int \Lambda_L Q_L^A(X) X^{\star}_A + \Lambda_R Q_R^A(X) X^{\star}_A
\end{equation}
The action $S$ is given by Eq. (\ref{NewAction}). Such a theory requires regularization.

The minimalistic sigma-model action is written in terms of the target space metric ${\bf G}$ 
and the B-field ${\bf B}$. For example, the action of Eq. (\ref{NewAction}) corresponds to:
\begin{align}
{\bf G}\;=\;& \mbox{Str}\left({1\over 2}J_2J_2 + J_1({\bf 1}-{\bf P}_{31}) J_3\right)
\\   
{\bf B}\;=\;& \mbox{Str}\left(
              {1\over 2}J_1\wedge J_3 - J_1\wedge{\bf P}_{31} J_3
              \right)
\end{align}
The existence of the $b$-ghost is equivalent to the metric being the Lie derivative along $Q$ of 
some symmetric tensor $\bf b$:
\begin{equation}
{\bf G} = {\cal L}_Q {\bf b}
\end{equation}
where ${\cal L}_Q$ is the Lie derivative along the vector field $Q$.
In our case (Appendix \ref{sec:BRSTofB}):
\begin{equation}
{\bf b}\;=\;
{\mbox{Tr}\left((\{J_3,\lambda_3\} - \{J_1,\lambda_1\}) J_2\right)
  \over
 \mbox{STr}(\lambda_3\lambda_1)}
\end{equation}
As in Section \ref{sec:B-ghost}, the part of the action involving the target space metric $\bf G$ is
BRST exact.

\section{Open problems}\label{sec:OpenProblems}
We did not verify that $\widehat{\Phi}'_{\xi}$ of Eq. (\ref{PhiWithConnection}) satisfies 
\amklink{math/bv/omega/Equivariant_Form.html}{the conditions}
formulated in \cite{Mikhailov:2016myt,Mikhailov:2016rkp}. In particular, we may hope for
$\{\widehat{\Phi}'_{\xi},\widehat{\Phi}'_{\xi}\}=0$, but more complicated scenarios are also possible.
We believe that the invariance of our construction under the symmetries of 
$AdS_5\times S^5$ is important to satisfy those conditions at the quantum level.

We did not explicitly calculate the restriction of the $\widehat{\Phi}'_{\xi}$ to the standard family of Lagrangian
submanifolds, corresponding to the integration over the space of metrics. It can probably be
expressed in terms of $\cal O$ where $\overline{\partial}b = Q{\cal O}$ as calculated in \cite{Berkovits:2010zz}. In any case, it is most 
likely nonzero, and therefore the string measure of Eq. (\ref{StringMeasure}) is not just the product of 
Beltrami differentials, but involves also the curvature terms $\widehat{\Phi}'_F$.

\section*{Acknowledgments}
We want to thank Nathan~Berkovits, Henrique~Flores and Renann~Lipinski for discussions
and comments. This  work  was supported in part by the 
FAPESP grant 2014/18634-9 ``Dualidade Gravitac$\!\!,\tilde{\rm a}$o/Teoria de Gauge''
and in part by the
RFBR grant 15-01-99504 ``String theory and integrable systems''. 

\appendix

\section{The projector}\label{sec:Projector}
\subsection{Definition}
Let $\pi_{A}$ and $\pi_S$ denote the projectors:
\begin{align}
&   \pi_A\;:\;   T(AdS_5\times S^5) \to T(AdS_5) \;
   \mbox{ \tt\small projector along }\; T(S^5)
   \\    
&   \pi_S\;:\;   T(AdS_5\times S^5) \to T(S^5) \;
   \mbox{ \tt\small projector along }\; T(AdS_5)
   \\  
&   \pi_A(v) + \pi_S(v) =\; v
\end{align}
\marginpar{\small over- \\line $\overline{v}$}
For any vector $v$, we will denote by $\overline{v}$ the difference of its $AdS_5$ and $S^5$ components:
\begin{equation}\label{DefinitionOfVBar}
   \overline{v} \;\;\stackrel{\tiny\rm def}{=}\;\; \pi_A(v) - \pi_S(v)
\end{equation}
The projector ${\bf P}_{13}\;:\;{\bf g}_1\to {\bf g}_1$ was defined in \cite{Bedoya:2010qz} as follows:\marginpar{${\bf P}_{13}$}
\begin{align}
   {\bf P}_{13} A_1 \;&= A_1 + [S_2,\lambda_3]
\label{ProjectorAlongS2Lambda3}
\\  
\{\lambda_1\;,\; {\bf P}_{13} A_1 \} \;&= 0
\label{ProjectorOnAnnLambda1}
\end{align}
where $S_2\in {\bf g}_2$ is adjusted to satisfy (\ref{ProjectorOnAnnLambda1}). In fact ${\bf P}_{13}$ is the projection
to the tangent space $TC_R$ along the space $T^{\perp}C_L$ which is orthogonal  to $TC_L$ with
respect to the metric defined by $\mbox{Str}$: \marginpar{$T^{\perp}C$}
\begin{equation}
({\bf 1} - {\bf P}_{13})A_1\in T^{\perp}C_L
\end{equation}
In other words, for generic $\lambda_3$ and $\lambda_1$ we have an exact sequence:
\begin{align}
& 0\longrightarrow T^{\perp}C_L \stackrel{i}{\longrightarrow} {\bf g}_1 
\stackrel{{\bf P}_{13}}{\longrightarrow} TC_R \longrightarrow 0
\end{align}
In Section \ref{sec:ExplicitProjector} we will give an explicit formula for ${\bf P}_{13}$ following \cite{Berkovits:2010zz}.

\subsection{Matrix language}
It turns out that computations can often be streamlined by thinking about elements of ${\bf g}$
literally as $4|4$-matrices. In fact $\bf g$ is a {\em factorspace} of $sl(4|4)$ modulo a subspace 
generated by the unit matrix. Therefore, when talking about a matrix corresponding to an
element of $\bf g$, we have to explain every time how we choose a representative. The ${\bf Z}_4$
grading of $psl(4|4)$ can be extended to $sl(4|4)$; the unit matrix has grade two. Therefore,
the ambiguity of adding a unit matrix only arises for representing elements of ${\bf g}_2$. To deal
with this problem, we introduce some notations. Given a matrix $X$ of grade two, we denote by
$X_{\tl}$ the corresponding traceless matrix:\marginpar{TL}
\begin{equation}
X_{\tl} = X - {\mbox{Tr}(X)\over 8} {\bf 1}
\end{equation}
(The subscript ``TL'' is an abbreviation for ``traceless''.)
Also, it is often useful to consider $4|4$-matrices with nonzero supertrace. Such matrices
do not correspond to any elements of $\bf g$. For a $4|4$-matrix $Y$ we define:\marginpar{STL}
\begin{align}
Y_{\stl} \;=\; & Y - {\mbox{STr}(Y)\over 8} \Sigma
\\     
\mbox{\tt\small where } & \Sigma = \mbox{diag}(1,1,1,1,-1,-1,-1,-1)
\end{align}
In particular:
\begin{equation}
(Y_{\tl})_{\stl} = (Y_{\stl})_{\tl} = Y - {\mbox{Tr}(Y)\over 8} {\bf 1} - {\mbox{STr}(Y)\over 8} \Sigma
\end{equation}
We also define, for any even matrix $Y$:\marginpar{$\Sigma$}
\begin{equation}
\overline{Y} = Y\Sigma = \Sigma Y
\end{equation}
This definition agrees with Eq. (\ref{DefinitionOfVBar}).

\subsection{Explicit formula for the projector}\label{sec:ExplicitProjector}
In fact $S_2$ is given by the following expression:
\begin{equation}\label{ExplicitExpressionForS2}
   S_2 = {2\over \mbox{Str}(\lambda_1\lambda_3)}\overline{\{\lambda_1,A_1\}}_{\stl}
\end{equation}
Notice that $\overline{\{\lambda_1,A_1\}}_{\stl}$ is actually both super-traceless {\em and} traceless; it is
the same as $\overline{\{\lambda_1,A_1\}_{\tl}}$ (with the overline extending over ``$\tl$'').
We have to prove that the $S_2$ defined this way satisfies (\ref{ProjectorOnAnnLambda1}). Indeed, we have:
\begin{align}
   [S_2,\lambda_3] \;&= {2\over \mbox{Str}(\lambda_1\lambda_3)}
\left[\overline{\{\lambda_1,A_1\}}_{\stl}\;,\;\lambda_3\right]
\end{align}
and we have to prove Eq. (\ref{ProjectorOnAnnLambda1}). We have:
\begin{align}
&     \left\{
      \lambda_1\;,\;\left[\overline{\{\lambda_1,A_1\}}_{\stl}\;,\;\lambda_3\right]
   \right\} \; =\;  \left\{
      \lambda_1\;,\;\left[\Sigma\{\lambda_1,A_1\}_{\rm TL}\;,\;\lambda_3\right]
   \right\} \; =
\nonumber\\   
=\;&
- \Sigma \left[
   \lambda_1\;,\;\left\{
      \{\lambda_1,A_1\}_{\rm TL},\lambda_3
   \right\}
\right] = -\Sigma \left\{
   [\lambda_1,\lambda_3]\;,\;\{\lambda_1,A_1\}_{\rm TL}
\right\}
\label{ProductOfMatrices}
\end{align}
Both $\{\lambda_1,A_1\}$ and  $[\lambda_1,\lambda_3]$ have ${\bf Z}_4$-grading two. Let us use:
\begin{equation}\label{CommutatorOfPureSpinors}
   [\lambda_1,\lambda_3] = {1\over 4}\mbox{Str}(\lambda_1\lambda_3)\Sigma + 
[\lambda_1,\lambda_3]_{\rm STL}
\end{equation}
For all grade $\bar 2$ matrices $A_2$ and $B_2$ such that 
$\mbox{Tr}A_2 = \mbox{Tr}B_2 = \mbox{STr}A_2 = \mbox{Str}B_2=0$ 
\amklink{math/pure-spinor-formalism/AdS5xS5/Symmetries.html}{the following identity holds}:
\begin{equation}
 \{A_2,B_2\} \;=\; A_2B_2 + B_2A_2 \;=\; {1\over 4}\left(\mbox{Str}(A_2B_2)\Sigma + \mbox{Tr}(A_2B_2){\bf 1}\right)
\end{equation}
Therefore:
\begin{align}
&     \left\{
      \lambda_1\;,\;\left[\overline{\{\lambda_1,A_1\}}\;,\;\lambda_3\right]
   \right\} \mbox{ \tt\small mod } {\bf 1}\; =\; 
 -{1\over 2}\mbox{Str}(\lambda_1\lambda_3) \{\lambda_1,A_1\}_{\rm TL}
\end{align}
(where ``$\mbox{\tt mod} {\bf 1}$'' means ``modulo the center of $psl(4|4)$'', {\it i.e.} up to a multiple of the
unit matrix). This proves (\ref{ProjectorOnAnnLambda1}).

The central part of $\left\{
      \lambda_1\;,\;\left[\overline{\{\lambda_1,A_1\}}\;,\;\lambda_3\right]
   \right\}$ is generally speaking nonzero:
\begin{align}
\mbox{Tr}
& \left\{
      \lambda_1\;,\;\left[\overline{\{\lambda_1,A_1\}}\;,\;\lambda_3\right]
   \right\}
\;=\;
2\mbox{Tr}\left(
      \lambda_1\left[\overline{\{\lambda_1,A_1\}}\;,\;\lambda_3\right]
\right)\;=\;
\\    
\;=\;&
-2\mbox{Tr}\left(
       [\lambda_1,\lambda_3]_{\rm STL}\Sigma\{\lambda_1,A_1\} 
\right)\;=\;
-2\mbox{STr}\left(
       [\lambda_1,\lambda_3]_{\rm STL}\;\{\lambda_1,A_1\} 
\right)
\end{align}
In $\Gamma$-matrix notations, $[\lambda_1,\lambda_3]_{\rm STL}$ is $(\lambda_1,\overline{\Gamma}^m\lambda_3)$ and $\{\lambda_1,A_1\}$ is $(\lambda_1,\Gamma^m A_1)$.

\vspace{10pt}\noindent
Let us define ({\it cp.} Eq. \ref{ExplicitExpressionForS2}):
\begin{align}
{\bf S}_{2/1}\;:\;& {\bf g}_1\rightarrow {\bf g}_2
\nonumber \\      
{\bf S}_{2/1}A_1\;=\;& {2\over \mbox{Str}(\lambda_1\lambda_3)}\overline{\{\lambda_1,A_1\}}_{\stl}
\label{S21}\\    
{\bf S}_{2/3}\;:\;& {\bf g}_3\rightarrow {\bf g}_2
\nonumber \\      
{\bf S}_{2/3}A_3\;=\;& {2\over \mbox{Str}(\lambda_3\lambda_1)}\overline{\{\lambda_3,A_3\}}_{\stl}
\label{S23}
\end{align}
so that:
\begin{align}
{\bf P}_{13}A_1 \;=\; & A_1 + [{\bf S}_{2/1}A_1, \lambda_3]
\\
{\bf P}_{31}A_3 \;=\; & A_3 + [{\bf S}_{2/3}A_3, \lambda_1]
\end{align}

\subsection{Properties of ${\bf P}_{13}$ and ${\bf P}_{31}$}
It follows from the definition, that for any $v_2\in {\bf g}_{\bar{2}}$ we have 
\begin{equation}\label{ProjectorOfV2Lambda3IsZero}
   {\bf P}_{13}[v_2,\lambda_3] = 0
\end{equation}
Let us verify this explicitly using the definition (\ref{ProjectorAlongS2Lambda3}) with the explicit expression 
for $S_2$ given by (\ref{ExplicitExpressionForS2}). We have:
\begin{align}\label{P13OnV2Lambda3}
   {\bf P}_{13} [v_2,\lambda_3] \;& = [v_2,\lambda_3] + 
   {2\over \mbox{Str}(\lambda_1\lambda_3)}
   \left[
      \;\overline{\{\lambda_1,[v_2,\lambda_3]\}}\;,\;\lambda_3\;
   \right]
\end{align}
Consider the expression $\left[\;\overline{\{\lambda_1,[v_2,\lambda_3]\}}\;,\;\lambda_3\;\right]$:
\begin{align}\label{Lambda1V2Lambda3Lambda3}
   \left[\;\overline{\{\lambda_1,[v_2,\lambda_3]\}}\;,\;\lambda_3\;\right] \; = \; &
   \left[\;\Sigma\{\lambda_1,[v_2,\lambda_3]\}\;,\;\lambda_3\;\right] \; = \; 
  \\ 
   \;=\; & -\left[\;\Sigma\{[\lambda_1,\lambda_3],v_2\}\;,\;\lambda_3\;\right] +
   \left[\;\Sigma[\{\lambda_1,v_2\},\lambda_3]\;,\;\lambda_3\;\right] 
\end{align}
Let us consider the first expression on the RHS of (\ref{Lambda1V2Lambda3Lambda3}). 
Using (\ref{CommutatorOfPureSpinors}) we rewrite:
\begin{align}
&   
     - \left[\;\Sigma \{[\lambda_1,\lambda_3],v_2\}\;,\;\lambda_3\;\right] \;=
\nonumber\\    
=\;&\;
- {1\over 4} \mbox{Str}(\lambda_1\lambda_3)[\Sigma \{\Sigma,v_2\}\;,\;\lambda_3]  \; =
\; - {1\over 2}\mbox{Str}(\lambda_1\lambda_3)\;[v_2,\lambda_3]
\end{align}
This cancels with the first term on the RHS of (\ref{P13OnV2Lambda3}). And the second expression on 
the RHS of (\ref{Lambda1V2Lambda3Lambda3}) is zero:
\begin{equation}
   \left[\;\Sigma [\{\lambda_1,v_2\},\lambda_3]\;,\;\lambda_3\right]\;=\;
\left[\; \{\Sigma\{\lambda_1,v_2\},\lambda_3\}\;,\;\lambda_3\;\right]\;=\;0
\end{equation}

\subsection{Subspaces of $\bf g$ associated to pure spinors}\label{sec:Subspaces}
Consider the decomposition:
\begin{align}
{\bf g}_2 \;=\; & {\bf g}_{2L} \oplus {\bf g}_{2R}
\oplus {\bf C} [\lambda_3,\lambda_1]_{\rm STL} 
\oplus {\bf C} \overline{[\lambda_3,\lambda_1]}_{\tl}
\end{align}
Here ${\bf g}_{2L}$ is a 4-dimensional subspace $\mbox{Tr}$-orthogonal to ${\bf C} \overline{[\lambda_3,\lambda_1]}_{\tl}$ and commuting with $\lambda_3$,
and ${\bf g}_{2R}$ is  $\mbox{Tr}$-orthogonal to ${\bf C} \overline{[\lambda_3,\lambda_1]}_{\tl}$ and commuting with $\lambda_1$.

Similarly we can refine $T^{\perp}C_R$  and $T^{\perp}C_L$:
\begin{align}
{\bf g}_3\supset T^{\perp}C_R \;=\;& [{\bf g}_{2L},\lambda_1] \oplus {\bf C}[[\lambda_3,\lambda_1]_{\rm STL},\lambda_1]
\\    
{\bf g}_1\supset T^{\perp}C_L \;=\;& [{\bf g}_{2R},\lambda_3] \oplus {\bf C}[[\lambda_3,\lambda_1]_{\rm STL},\lambda_3]
\end{align}

\section{BRST variation of the $b$-tensor}\label{sec:BRSTofB}
Here we will prove:
\begin{equation}
(Q_L + Q_R) \frac{\mbox{Tr}\{J_1,\lambda_1\}J_2}{\mbox{Str}(\lambda_3\lambda_1)} \;=\;
-\mbox{Str}\left({1\over 4}J_2J_2 + {1\over 2}J_1({\bf 1}- {\bf P}_{31})J_3 \right)
\end{equation}
(Remember that $\mbox{Tr}\,\ldots\, = \mbox{Str}(\ldots\Sigma)$.)
In fact, only $Q_L$ contributes; the action of $Q_R$ is zero:
\begin{equation}
Q_R\;\mbox{Str}\left(\{J_1,\lambda_1\}J_2\Sigma\right)\;=\; - \mbox{Tr}\left( \{J_1,\lambda_1\}\{J_1,\lambda_1\}_{\rm TL}\right)\;=\;0
\end{equation}
because $J_1$ is a fermion. Let us compute the action of $Q_L$:
\begin{align}
&Q_L\;\mbox{Str}\left(\{J_1,\lambda_1\}J_2\Sigma\right)= 
- \mbox{Str}\left(\left( \{[J_2, \lambda_3],\lambda_1\}J_2 + 
\{J_1,\lambda_1\}\{J_3,\lambda_3\}_{\rm TL}\right)\Sigma\right)=
\nonumber\\    
\;=\;
& - \mbox{Tr}\left(\{J_2,[J_2,\lambda_3]\}\lambda_1\right) - 
\mbox{Str}\left(\overline{\{J_1,\lambda_1\}}_{\rm STL}\{J_3,\lambda_3\}\right)\;=
\nonumber\\  
\;=\;
& - {1\over 4}\mbox{Str}(\lambda_3\lambda_1)\mbox{Str}(J_2^2) -
\mbox{Str}\left( J_3\left[\overline{\{J_1,\lambda_1\}}_{\rm STL},\lambda_3\right]\right)\;=
\nonumber\\   
\;=\;
& - {1\over 4}\mbox{Str}(\lambda_3\lambda_1)\mbox{Str}(J_2^2) -
  {1\over 2}\mbox{Str}(\lambda_3\lambda_1)\mbox{Str}\left(J_3({\bf 1}-{\bf P}_{13})J_1\right)
\end{align}


\begin{thebibliography}{10}

\bibitem{Berkovits:2010zz}
N.~Berkovits and L.~Mazzucato, {\it {Taming the b antighost with Ramond-Ramond
  flux}},  {\em JHEP} {\bf 11} (2010) 019 doi: {\bf 10.1007/JHEP11(2010)019}
  [{\tt arXiv/1004.5140}].

\bibitem{Mikhailov:2016myt}
A.~Mikhailov and A.~Schwarz, {\it {Families of gauge conditions in BV
  formalism}},  {\em JHEP} {\bf 07} (2017) 063 doi: {\bf
  10.1007/JHEP07(2017)063} [{\tt arXiv/1610.02996}].

\bibitem{Mikhailov:2016rkp}
A.~Mikhailov, {\it {Integration over families of Lagrangian submanifolds in BV
  formalism}},  {\em Nucl. Phys.} {\bf B928} (2018) 107--159 doi: {\bf
  10.1016/j.nuclphysb.2018.01.006} [{\tt arXiv/1611.04978}].

\bibitem{Berkovits:2008ga}
N.~Berkovits, {\it {Simplifying and Extending the AdS(5) x S**5 Pure Spinor
  Formalism}},  {\em JHEP} {\bf 09} (2009) 051 doi: {\bf
  10.1088/1126-6708/2009/09/051} [{\tt arXiv/0812.5074}].

\bibitem{Tonin:2013uec}
M.~Tonin, {\it {On Semiclassical Equivalence of Green-Schwarz and Pure Spinor
  Strings in AdS(5) x S(5)}},  {\em J. Phys.} {\bf A46} (2013) 245401 doi: {\bf
  10.1088/1751-8113/46/24/245401} [{\tt arXiv/1302.2488}].

\bibitem{WithRenann}
R.~Lipinski, A.~Mikhailov, to appear

\bibitem{Berkovits:2000fe}
N.~Berkovits, {\it {Super Poincare covariant quantization of the superstring}},
   {\em JHEP} {\bf 04} (2000) 018 doi: {\bf 10.1088/1126-6708/2000/04/018}
  [{\tt arXiv/hep-th/0001035}].

\bibitem{Berkovits:2007rj}
N.~Berkovits and C.~Vafa, {\it {Towards a Worldsheet Derivation of the
  Maldacena Conjecture}},  {\em JHEP} {\bf 03} (2008) 031 doi: {\bf
  10.1088/1126-6708/2008/03/031} [{\tt arXiv/0711.1799}].

\bibitem{Alexandrov:1995kv}
M.~Alexandrov, M.~Kontsevich, A.~Schwartz, and O.~Zaboronsky, {\it {The
  Geometry of the master equation and topological quantum field theory}},  {\em
  Int. J. Mod. Phys.} {\bf A12} (1997) 1405--1430 doi: {\bf
  10.1142/S0217751X97001031} [{\tt arXiv/hep-th/9502010}].

\bibitem{Bedoya:2010qz}
O.~A. Bedoya, L.~Bevilaqua, A.~Mikhailov, and V.~O. Rivelles, {\it {Notes on
  beta-deformations of the pure spinor superstring in AdS(5) x S(5)}},  {\em
  Nucl.Phys.} {\bf B848} (2011) 155--215 doi: {\bf
  10.1016/j.nuclphysb.2011.02.012} [{\tt arXiv/1005.0049}].

\bibitem{Apfeldorf:1994av}
K.~M. Apfeldorf and C.~Ordonez, {\it {Field redefinition invariance in quantum
  field theory}},  {\em Nucl. Phys.} {\bf B479} (1996) 515--526 doi: {\bf
  10.1016/0550-3213(96)00451-8} [{\tt arXiv/hep-th/9408100}].

\bibitem{Kreimer:2012qu}
D.~Kreimer and A.~Velenich, {\it {Field diffeomorphisms and the algebraic
  structure of perturbative expansions}},  {\em Lett. Math. Phys.} {\bf 103}
  (2013) 171--181 doi: {\bf 10.1007/s11005-012-0589-y} [{\tt arXiv/1204.3790}].

\end{thebibliography}

\def\cprime{$'$} \def\cprime{$'$}
\providecommand{\href}[2]{#2}\begingroup\raggedright\endgroup

\end{document}